\documentclass{ar-1col}

\usepackage[letterpaper]{geometry}
\jname{Xxxx. Xxx. Xxx. Xxx.}
\jvol{AA}
\jyear{YYYY}
\doi{10.1146/((please add article doi))}

\usepackage{subcaption}
\usepackage{natbib}
\usepackage{xcolor}
\usepackage{amssymb}

\def\msun{\hbox{M$_\odot$}}

\def\fenrich{\hbox{f$_{\rm enriched}$}}

\begin{document}


\markboth{Bastian \& Lardo}{Multiple Populations}

\title{Multiple Stellar Populations in Globular Clusters}

\author{Nate Bastian$^1$ and Carmela Lardo$^{1,2}$
\affil{$^1$Astrophysics Research Institute, Liverpool John Moores University, 146 Brownlow Hill, Liverpool, UK L3 5RF; email: N.J.Bastian@ljmu.ac.uk\\
$^2$ Laboratoire d'astrophysique, \'Ecole Polytechnique F\'ed\'erale de Lausanne (EPFL), Observatoire, 1290, Versoix, Switzerland; email: carmela.lardo@epfl.ch}
}

\begin{abstract}
Globular Clusters (GCs) exhibit star-to-star variations in specific elements (e.g., He, C, N, O, Na, Al) that bare the hallmark of high temperature H burning.  These abundance variations can be observed spectroscopically and also photometrically, with the appropriate choice of filters, due to the changing of spectral features within the band pass. This phenomenon is observed in nearly all of the ancient GCs, although, to date, has not been found in any massive cluster younger than $2$~Gyr.  Many scenarios have been suggested to explain this phenomenon, with most invoking multiple epochs of star-formation within the cluster, however all have failed to reproduce various key observations, in particular when a global view of the GC population is taken.  We review the state of current observations, and outline the successes and failures of each of the main proposed models.  The traditional idea of using the stellar ejecta from a 1st generation of stars to form a 2nd generation of stars, while conceptually straight forward, has failed to reproduce an increasing number of observational constraints.  We conclude that the puzzle of multiple populations remains unsolved, hence alternative theories are needed.

\end{abstract}

\begin{keywords}
keywords, separated by comma, no full stop, lowercase
\end{keywords}
\maketitle

\tableofcontents

\section{Introduction}

The traditional concept of globular clusters (GCs) as simple stellar populations, where all stars share the same age and abundances within some small tolerance, is now a view of the past, as it has become clear that (nearly) all GCs host significant abundance spreads within them.  While all GCs show the same basic pattern, enriched populations in He, N, \& Na and populations depleted in O \& C, the specifics of each cluster are unique.  It is the manifestations of these distinctive chemical anomalies that cause the impressively complex colour-magnitude diagrams (CMDs) that have been uncovered with precision Hubble Space Telescope (HST) photometry, especially when viewed in the UV and near-UV. These star-to-star abundance variations within clusters are known as ``multiple populations" (MPs).

\begin{marginnote}[]
\entry{GCs}{Globular Clusters}
\entry{MPs}{Multiple Populations}
\end{marginnote}

The past decade has seen an impressive amount of observational work on the topic, with ground based spectroscopic surveys of thousands of stars within samples of GCs tracing the detailed abundance patterns \citep[e.g.,][]{Carretta:09UVES}, and space-based photometry providing unprecedented views of the number and make-up of the different populations within the GCs \citep[e.g.][]{Piotto:15UVsurvey}.  In addition to these observational advances, a number of scenarios for the origin of MPs have been put forward, which have begun providing testable predictions.  Alongside the co-formation/evolution of GC populations in galaxies, the origin of MPs is one of the major unsolved problems in GC and stellar populations research.

The goal of this review is to provide an overview of the present state of observations  of MPs along with a critical comparison against theoretical models that have been put forward for their origin.  We focus the majority of our attention to results obtained since the last {\em Annual Review} on the topic \citep{Gratton:04} and refer the interested reader to that comprehensive review for the historical developments and status of the field up until that time.  Additionally, there have been a number of more recent excellent reviews on the topic, notably \citet{Gratton:12Rev} and \citet{Charbonnel:16}.  The field has been growing at a rapid rate, with hundreds of relevant papers published each year, and as such, we are unable to reference all work in the field.  Instead we use typical examples to illustrate broader points, and attempt to synthesise all results into a coherent status update of the field.

While many of the previous reviews have concentrated on the chemistry of MPs, we explore that as only one line of evidence, and also consider global properties and correlations, relation to field stars and the physical properties of both young and old massive clusters.

We define MPs as the presence of star-to-star variations in chemical abundances, not expected from stellar evolutionary processes.  In particular, as will be reviewed below, this means variations in light elements such as He, C, N, and O that can cause complexities in CMDs as well as Na, Al and in some cases Mg. This can be contrasted with observations of some young ($<2$~Gyr) clusters which show unexpected features in their CMDs (e.g., extended main sequence turn-offs or split main sequences) which are not due by abundance variations but are rather driven by stellar evolutionary processes (i.e., rotationally induced stellar structure changes).

Finally, a note about terminology.  Stars within GCs that show enhancements in He, N, Na and are depleted in O and C have various labels in the literature, e.g., ``2nd generation stars", ``2nd population", ``enriched stars"\footnote{We use to term ``enriched" in the ``chemical enrichment" sense, meaning that the material appears to be processed through nuclear reactions in stars.   We note that some elements are in fact depleted (e.g., O, C).}.  We chose to use the latter options, as ``2nd generation" implies a genetic link to a first population.  While such a link is possible, it is by no means established (in fact evidence is currently pointing away from this interpretation), hence the use of more neutral terminology is more natural as the origin of MPs is still unknown.  However, when referring to models that explicitly invoke multiple generations of stars, we will use the ``generation" label for clarity. Also, we will use ``correlation" to refer to a positive correlation between two or more elements, and ``anti-correlation" for a negative correlation between abundances.

\section{Observations of Abundance Variations and Colour Magnitude Diagrams}

\subsection{Abundance variations}
MPs with distinctive light element abundance pattern are widely observed in old  and massive clusters. Abundance spreads are only rarely associated with star-to-star Fe and heavy element variations, implying that  some unique chemical enrichment mechanism, operating only in cluster environments, is responsible for the observed chemistry. The suggestion that the light element anomalies arise from nuclear processing within massive stars from a previous generation born within GCs still remains the only theory that has been {\em quantitatively} investigated. Nonetheless, such a hypothesis suffers from several drawbacks and can only account for some of the relevant observations. In the following we will review the status of observations and critically discuss their interpretation in the framework of MPs.

\subsubsection{Light element abundance spreads}\label{SEC:CN}
The presence of chemical inhomogeneities among bright giants in clusters was revealed by pivotal studies in the early seventies \citep[e.g.][]{Osborn:71}. Stars at the same magnitude along the RGB were found to display variations in the strengths of CH, CN, and NH blue absorption features, due to underlying star-to-star variations in C and N abundances \citep{Bell:1980}\footnote{
In a first approximation, the CH absorption at 4300~\AA~can be regarded as a C sensitive diagnostic, while the CN and NH band-strengths (at 3839 and 3360\AA~respectively) are proxies for N.}.  Most of the studied GCs display either a bimodal or multimodal CN distribution \citep[e.g.][]{Norris:87}. The molecular CN (NH) and CH bands were found also to be anti-correlated, with CN-strong stars also characterised by weak CH absorption and vice versa; i.e. N is found to anti-correlate with C.

\begin{marginnote}[]
\entry{Primordial star (1P)}{star having the same abundances as the field at the same metallicity [Fe/H].}
\entry{Enriched stars (2P)}{star showing enhanced N, Na, and Al and depleted C and O abundances
with respect to field at the same metallicity [Fe/H].}
\end{marginnote}

While extremely common in clusters, stars characterised by enhanced N and depleted C are rarely found in the field and not present in open clusters \citep[OCs; e.g.][]{MacLean:15,Martell:11}.  However, GCs also contain stars that are characterised by the same abundance pattern observed in field stars of the same metallicity.  This has led to the notion that GCs are made up of MPs, one with field-like composition, and a second with ``anomalous chemistry" unique to GCs.  In the following, we will refer to the stars with peculiar chemical composition as enriched or 2P (second population) and the stars having field-like abundances as primordial or 1P (first population). We consider enriched or 2P and primordial  or 1P as synonyms and we use the expressions interchangeably throughout this review.

Evolutionary mixing was originally proposed as the main cause of the C and N inhomogeneities as normal stellar evolution may contribute to the observed N-C anti-correlation in evolved RGBs \citep[e.g.,][]{Deni:90}. However, such an {\em evolutionary} scenario was soon challenged by observations \citep[e.g.][]{Gratton:04},  as 
mixing theories cannot explain the abundance anomalies seen among  non-evolved or scarcely evolved MS and MSTO stars \citep[e.g.][]{Cannon:98,Briley:04} which are characterised by negligible outer convective zones. Even if sufficient mixing could be achieved during MS evolution, it would also result in changes in helium abundances and extended lifetime of stars, e.g. mixing would result in broadening the MSTO region in the CMD, contrary to what observed (in ancient GCs).

\begin{marginnote}[]
\entry{RGB}{Red Giant Branch}
\entry{HB}{Horizontal Branch}
\entry{AGB}{Asymptotic Giant Branch}
\entry{SGB}{Sub-Giant Branch}
\entry{MS}{Main Sequence}
\entry{MSTO}{Main Sequence Turn-off}
\end{marginnote}

\begin{figure}[!b]
    \centering
    \begin{minipage}{.4\textwidth}
        \centering
        \includegraphics[width=1\linewidth]{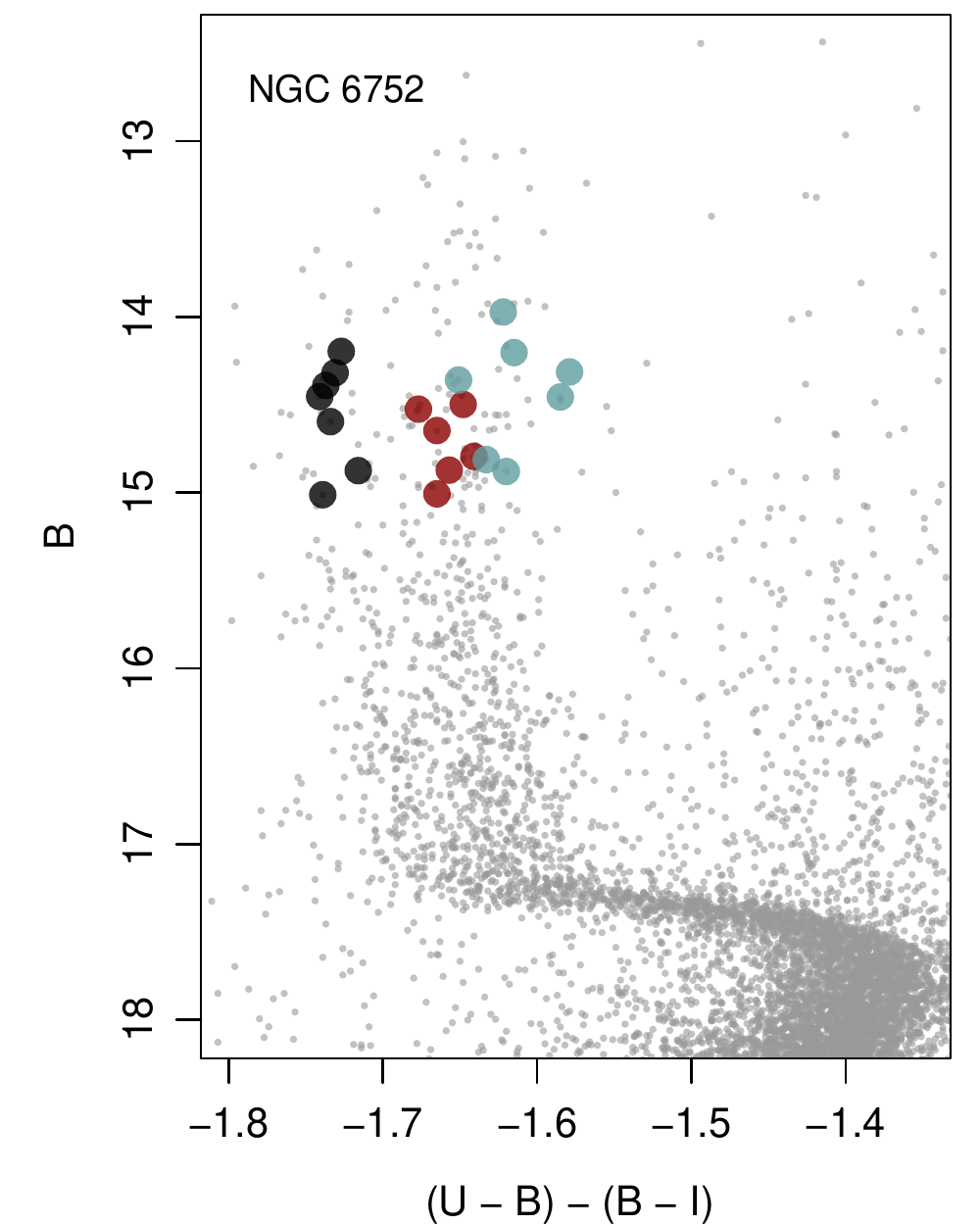}

    \end{minipage}%
    \begin{minipage}{0.6\textwidth}
        \centering
        \includegraphics[width=1.\linewidth]{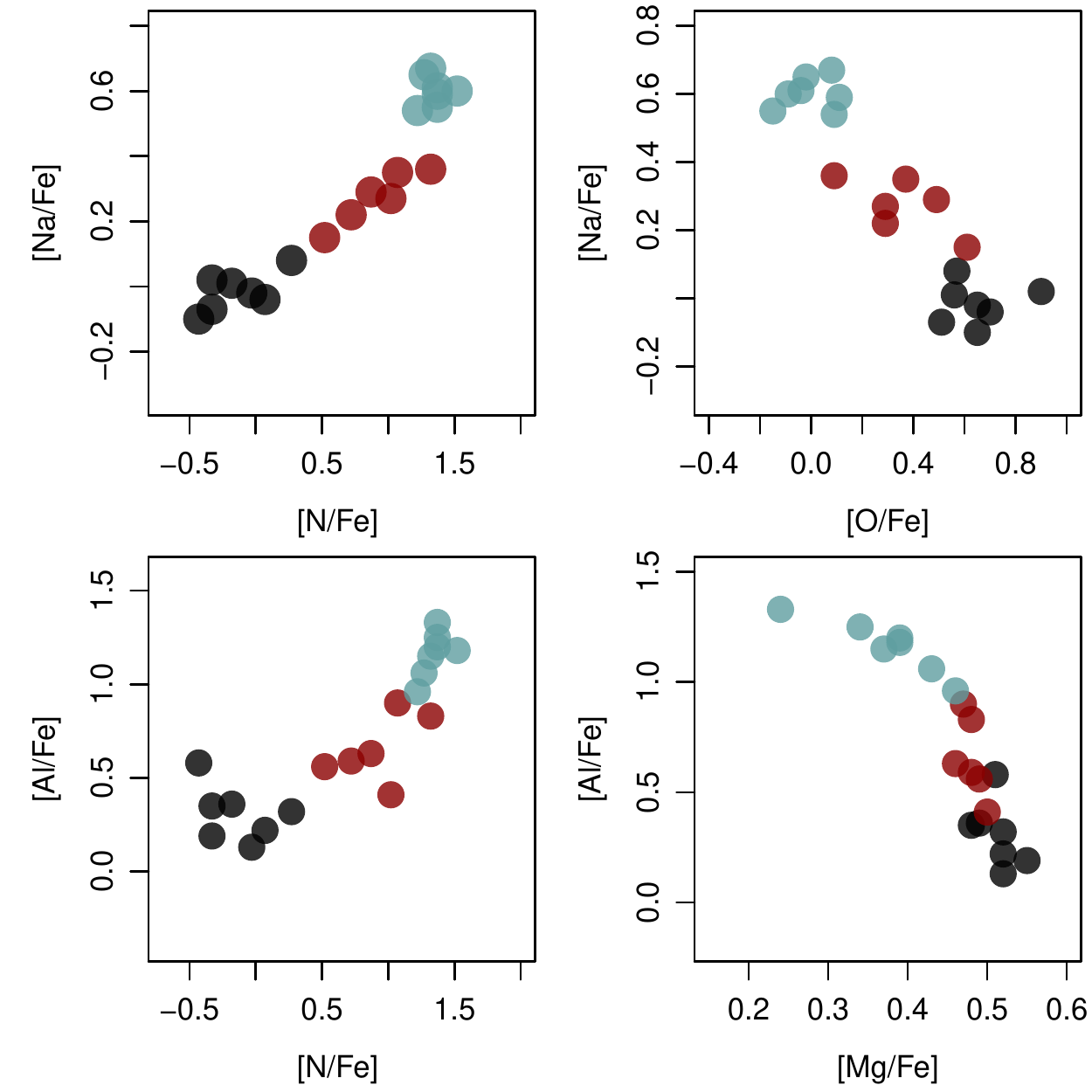}

    \end{minipage}
 \caption{NGC~6752 C$_{\rm U,B,I}$ vs. V CMD is shown in the {\bf left panel}.  Photometry has been kindly provided by Peter Stetson. Spectroscopic targets from \citet{Yong:05,Yong:08} are also plotted. Colours correspond to a different chemical composition, with green, red, and black symbols having high, moderate, and primordial Na content. Stars with different light-element composition which are well mixed along the RGB in optical colours, occupy 
 distinct sequences in the C$_{\rm U,B,I}$ vs. V CMD. The same stars are plotted in the {\bf middle} and {\bf right panels} to show light abundance variations.}
\label{fig:CUBI}
   
\end{figure}

When higher-resolution spectra allowed for direct spectroscopic measurements of Na and O (through atomic lines) in
 stars where N and C abundances were available, it was found that the N overabundance (C depletion) was associated to enhanced Na (O depletion); i.e. N-Na and C-O are positively correlated \citep[e.g.][]{Sneden:92}. Also, while the individual abundances of C, N, O show large spreads, the sum C+N+O is  generally observed to be constant  \citep[e.g.][see also~\S~\ref{sec:SCLUSTERS}]{Dickens:91}.
Anti-correlated Na and O ranges were found in nearly all the studied clusters, along with variations in Al and (possibly) Mg, anti-correlated with each other \citep[e.g.][]{Gratton:04,Gratton:12Rev}.
 While O can potentially be depleted in the interiors of low mass stars through the CNO-cycle reactions, 
variations in the abundances of heavier elements like Na, Al, and Mg cannot by produced by 
fusion reactions within low-mass stars. This is because their temperatures are too low for the p-capture reactions to operate through the NeNa- and MgAl-chains (e.g.; \citealp{Prantzos:07}, Prantzos, Iliadis \& Charbonnel 2017). Hence, the abundance anomalies are not produced in the course of the evolution of stars we are currently observing but they were produced elsewhere, potentially within the interiors of more massive stars. See Fig.~\ref{fig:CUBI} (right panels) for some of the typical (anti-)correlations associated with MPs.

\begin{marginnote}[]
\entry{FRMS}{Fast Rotating Massive Star}
\entry{VMS}{Very Massive Star ($\geq$5000~\msun)}

\end{marginnote}

How this material would then find its way into the low mass stars observed today is still an open question, as is the exact source of the material.  Most models to date have adopted a scenario where material from a first generation of stars  pollutes the intra cluster medium out of which subsequent generations of stars were born.  Several candidate 1P polluters -- either intermediate mass asymptotic giant branch stars (AGBs; $3-9$~\msun), fast rotating massive  (FRMSs, $>15$~\msun), or  very massive (VMSs; $\geq$5000~\msun) stars -- have been proposed because they are
sites of hot CNO and NeNa processing and we will discuss them in \S~\ref{SEC:NUCLEARREACTION}.
A (weak) Si-Mg anti-correlation was observed in a small number of massive and/or metal-poor GCs \citep[e.g. NGC~6752, NGC~2808, M~15;][]{Yong:05,Carretta:09UVES}, implying proton burning occurring in even hotter environments ($\geq$75 MK), than needed for the CNO and NeNa processing. 

The presence of anti-correlated CNONaAl abundances has been demonstrated to be nearly universal among old and massive clusters and has even been suggested to be {\em the} distinguishing feature between {\em genuine} globulars from other stellar associations \citep[e.g., OCs or dwarf galaxies;][]{Carretta:10GLOB}.  If stars with high N, Na, or Al abundances are found in the field, they are usually considered to have originated from GCs (unless they are part of a binary system).  Spectroscopic studies have estimated that  at least 3\% of the local field metal-poor star population was born in GCs \citep[e.g.,][]{Carretta:10GLOB,Martell:11}, under the assumption that all 2P stars must form in GCs.

The shape and the extension of the light element anti-correlations (i.e. their extrema, substructure and their multi-modality) vary from cluster to cluster, with some clusters showing a well extended Na-O anti-correlation and objects for which both Na and O abundances span very short 
ranges \citep[e.g.][]{Carretta:09GIR,Carretta:09UVES}. 
 In a few cases, the Na-O distribution is clumpy, with the presence of one or more gaps
\citep[e.g.][]{Marino:08M4, Lind:11,Carretta:15N2808}. Such quantised distributions may be common, but measurements with very small uncertainties are needed to corroborate this. However, such a multimodality of the blue CN band is nearly universal in metal-rich clusters ([Fe/H] $\geq$ --1.7 dex) where errors on CN measurements are small enough to reveal discrete distributions \citep[][]{Norris:87}. 

The light element variations span similar intervals in different evolutionary phases \citep[e.g.][]{Gratton:12HBN1851}.
Observations show that unevolved stars on the MS and evolved RGB stars span the same ranges of chemical anomalies demonstrate that such light element variations cannot be due to accretion of processed material on already formed stars, as the anti-correlations would be strongly diluted by mixing as the stars evolve \citep[e.g.][]{Gratton:04}. Also, the ratio between 1P and 2P stars along the AGB appears to be consistent with
the corresponding ratio found on the RGB and the observed HB morphology  \citep[e.g.][]{Cassisi:14,Lapenna:16,Lardo:17}.

An Al-Mg anti-correlation is not observed in all the GCs where the Na-O and N-C variations are detected. There are clusters that are characterised by a single Al abundance, while others show wide Al ranges \citep[][]{Carretta:09UVES,Meszaros:15}.  The majority of the MW GC stars for which Mg abundances are available have typical Mg abundances in the range 0.2 $\leq$ [Mg/Fe] $\leq$ 0.5 dex; implying a very short (if any) Al-Mg anti-correlation. Only a few Galactic GCs have been found to host stars that are  significantly deficient in Mg ([Mg/Fe]$\leq$ 0.0 dex -  \citealp[e.g.][]{Carretta:142808,Mucciarelli:12N2419}).
The extent of the Al-Mg anti-correlation correlates with both cluster mass and metallicity, as massive and metal-poor cluster tend to have larger Al-Mg anti-correlations \citep[e.g.][]{Carretta:09UVES,Carretta:09GIR,Pancino:17}.

While the N-C, Na-O (and in some cases the Al-Mg) anti-correlations and photometric spreads along the RGBs (see~\S~\ref{SEC:PHOTOMETRY}) are distinctive signatures present in (nearly) all ancient GCs, the cluster-to-cluster differences are large in terms of the extreme values, substructure and multi-modality. The evidence that each surveyed GC has its own specific pattern of MPs calls for a high degree of variety (or stochasticity) that must be taken into account when proposing MP formation mechanisms \citep[e.g.,][]{BastianHe}.
  
To date, there have only been a few stars in a handful of GCs that have been fully characterised in terms of their chemistry  \citep[i.e. the full set of varying elements: C, N, O, Na, Al, Mg, He, s-process, etc; e.g.][]{Smith:15}.  Instead, different surveys have focussed on different elements, and often even different stars within the same GCs.  This is an obvious avenue for future studies, to characterise the exact chemical fingerprint of 1P and enriched 2P stars. We refer the interested reader to the compilation of \citet[][]{Roediger:14} for abundances for a number of elements for stars in GCs.

\subsubsection{He variations: Main Sequence splitting, Horizontal Branch morphology, direct measurements}\label{SEC:HE}

If the CNONaAlMg star-to-star variations are built in the stellar interiors through CNO cycling
and p-capture processes at high temperatures, we may also expect He variations (as it is the main product of H burning). 
The observational data suggest that N and Na variations are always correlated with some (variable) He enhancement. However, this result is mostly based on indirect evidence as only a handful of studies have provided direct He abundance determinations\footnote{Direct measurements of  elements like He, O, Na, and Al, require high-resolution thus they are limited to the brighter stars in GCs. Conversely, both N and C are generally measured in fainter stars at the base of the RGB, because evolutionary mixing as the star evolves along the RGB can blur the MP chemical signature. Hence abundance determinations for the whole set of elements which vary in GCs are available only for a few stars in a handful of clusters.}.

He enhancement can be inferred from:
{\em (a)} {\em direct} measurements of He abundances, {\em (b)} splits or spreads of the MS in optical CMDs, and {\em (c)}  the HB morphology of the clusters. In what follows we refer to the He mass fraction as Y and denote variations in He as 
$\Delta$Y = Y -- Y$_{\rm p}$, where Y$_{\rm p}$ represents the initial He mass fraction value of Y$_{\rm p}$ = 0.244
\citep[e.g.;][]{cassisi03}.

Direct Y measurements are difficult to obtain. Temperatures above T$>$ 8500 K are necessary to detect the He photospheric transitions in the optical. However, hot HB stars -- where the He line might appear because of their high temperatures, are also affected by diffusion and preferential settling of elements \citep{Behr:03}.
As a result, Y can only be measured in stars with temperatures between $\sim$8500 - 11500~K, which are hot enough to show He line but  still cooler than the Grundahl-jump \citep[e.g.;][]{moehler14}, the temperature limit above which the original surface abundances are changed by diffusion  \citep{Grundahl:99}. Nonetheless, Y measurements from the photospheric He I line at 5875~\AA~in HB stars have been obtained for some GCs, and variations have been reported, with typical spreads of $\Delta(Y)=0.02-0.05$ \citep[see][for a summary]{MucciarelliHe}. He-rich stars have been also shown to be Na-rich and they are systematically located towards the blue regions of the HB \citep[][]{Villanova:09}.

For FGK-type stars, no photospheric lines exist and He can only be measured from the 
purely chromospheric He I absorption line at 10830~\AA. 
Studies based on this near-infrared transition confirm that He enrichment generally correlates with [Na/Fe] and [Al/Fe].
\citet{Dupree:13} directly measured He abundances from the 10830~\AA~line in two giants in $\omega$~Cen. They estimate an He abundance of Y $\leq$ 0.22 (below the big bang nucleosynthesis value) for the 1P star and Y = 0.39 - 0.44 for the 2P one, with the He-rich star also enhanced in Al.
Similarly,  \citet{Pasquini:11} performed a differential analysis between two giant stars of NGC 2808 with different Na abundances. They estimated that the 2P star is more He enriched than the Na-poor one by $\Delta$Y =0.17. 

While the direct spectroscopic evidence of He-enhancement is somewhat sparse, several
 photometric studies provided evidence that such He variations are in place \citep[e.g.,][]{Maeder:06,Anderson:0947Tuc,Bellini:13,Nardiello:15}.
Photometric estimates of $\Delta$Y can be derived by assuming that the observed colour dispersions at a given magnitude on the MS in optical colours (i.e. V--I) are due primarily to He spreads. 
The measure of $\Delta$Y spreads from MS isochrone fitting presently appears the most reliable method to infer He dispersions (see \citealt{Cassisi:17} and  \S~\ref{SEC:PHOTOMETRY}) and recent results from HST photometry reveal that the observed He spreads $\Delta Y$ strongly correlates with present-day cluster mass and luminosity, with more massive clusters having larger He spreads \citep[e.g.][which will be discussed in detail in \S~\ref{sec:age_mass}]{Milone:15M62}.

In $\omega$~Cen, the presence of a split MS \citep[e.g.][]{Bellini:10}
has been interpreted in terms of a large variation in the abundance of helium \citep[$\Delta$Y $\sim$ 0.15; e.g.][]{King:12}. The observation that the bluer MS is also $\simeq$ 0.3 dex more metal-rich than the redder MS further supports the existence of such large He enhancement, as canonical stellar models would predict the bluer MS to be more metal-poor than the red one and only a high He value can explain the colour difference between the two MSs \citep{Piotto:05}.

 Large He variations are also observed in clusters with homogeneous iron content, as in NGC~2808 where three distinct MSs can be clearly identified in optical CMDs \citep[e.g.][]{Piotto:07}. Given the lack of an iron spread \citep[e.g.][]{Carretta:06}, the MS split is interpreted as being due to three groups of stars with different He \citep{Milone2808He} which are likely linked to the multimodal HB structure \citep{Dantona:05,Dalessandro:10} and the three chemically distinct groups observed along the RGB \citep{Carretta:142808}.  In NGC~2808, such He variations are also correlated with light element abundance spreads, in the sense that stars with 1P composition are associated to the red MS with primordial He content, while stars with high N, Na, and Al are located onto the He-rich, blue MS \citep{Bragaglia:10b}. 
 
  Variations of He between 1P and 2P stellar groups may also affect the colour and luminosity of the RGB bump; as shown in \citet{Bragaglia:10a}. 
 
Variations in the abundance of He can have a significant impact the HB morphology \citep{Rood:73,Dantona:02}. This is because He-rich stars evolve faster than those with primordial He and thus, at a given age, He-rich stars at  the MSTO are less massive \citep[e.g.,][]{Chantereau:16}. Hence, if both He-rich and He-poor stars experience the same mass loss during RGB evolution, they  should end up on the HB stars with different masses; i.e. different colours (see also \citealp{Norris6752}). Indeed, the HB morphology of several clusters has been modelled in terms of variable He \citep[e.g.][]{Caloi:07,Cassisi:09,Dantona:10,Dalessandro:13,DiCriscienzo:15}.

 Since He affects the HB morphology both in terms of temperature (due to mass-loss) and luminosity (because of the different contribution to the luminosity of the H-burning shell), variations in colour (e.g.; temperature) along the HB are largely degenerate with mass-loss and age. Interestingly, the presence of He-enhanced populations along the blue part of the HB can be inferred without making assumptions about the RGB mass loss when a combination of optical and far-UV magnitudes is used \citep[e.g.][]{Dalessandro:10,Dalessandro:13}.

Further spectroscopic evidence (not including the measurement of He abundances) strengthens the connection between the HB morphology and the chemical composition  \citep[e.g.,][]{Gratton:14,Lovisi:12,Schaeuble:15}.  For example, the extension of the Na-O anti-correlation correlates with the maximum temperature of stars along the HB, indicating that the same physical mechanism responsible for the extreme Na enhancement and O depletion is also responsible for the morphology of the blue tail at the at end of the 
HB sequence \citep[][]{Carretta:10GLOB}. This correlation is interpreted as an evidence that the HB morphology is determined not only by age and metallicity but also by the He abundance, as Na-rich stars are also He-rich \citep[e.g.][]{Gratton:10HB}. More massive clusters also tend to have HBs that are more extended towards higher temperatures \citep[][]{RecioBlanco:06}. This evidence in turn would again suggest that very massive GCs show larger extents of processing, i.e. very low O and high Na (see \S~\ref{sec:age_mass}).

\subsubsection{Lithium variations among GC stars}\label{SEC:LI}

Lithium traces mixing processes, as it is rapidly destroyed in p-captures at temperatures exceeding $\sim$ 2.5 MK. Thus, if high values of N, Na, and Al are produced through hot H-burning, 2P stars should be depleted in Li.
Some studies have revealed an anti-correlation between Na and Li, as expected \citep{PasquiniLi,Lind:09Li,Dorazi:15Li}.  However, importantly, other works have not found evidence for Li variations among stars with 1P and 2P composition \citep[e.g.][]{Mucciarelli:11Li}.   Since, Li is destroyed at relatively low temperatures (i.e., well below temperatures where Na is formed) any material that is enriched in Na should be Li free. In order to explain the presence of some Li in 2P stars, it has been suggested that the polluters' ejecta (i.e. Li free, Na, N-rich) must be mixed with unprocessed material; i.e. gas which has always been kept cooler than $\sim$ 2.5 MK \citep[][]{Prantzos:06}.  Such models are known as ``dilution models", see \S~\ref{sec:models}, \S~\ref{sec:dilution}, and Fig.~\ref{fig:dilution}.

AGBs can potentially produce Li through the \citet{Cameron:71} mechanism at the beginning of the hot bottom burning \citep[HBB; e.g ][]{Ventura:02}. However, the finding of exactly the same Li abundance (or barely different) between 1P and 2P stars indicates that if AGB stars were responsible for the observed anomalies, they must have been able to {\em (a)} produce the same amount of Li previously destroyed by nuclear burning and  {\em (b)} give yields close to the values of primordial nucleosynthesis.
This concurrence certainly requires a high degree of fine-tuning and thus this explanation is unsatisfactory. On the other hand, both massive and very massive star models requires mixing with pristine material to account for the presence of lithium in 2P stars because their ejecta are Li free. Thus, the maximum depletion of oxygen in the final enriched composition cannot exceed the depletion of Li \citep[][]{Salaris:14}\footnote{As the processed material is expected to be Li-free, whereas it is only depleted in O.} contrary to what is observed \citep[][]{Shen:10}.
As a matter of fact, all the proposed scenarios have major problems in reproducing the Li content observed in clusters, where small (or no) variations of Li are found associated with large variations of other light elements.

\subsubsection{Mg \& K}\label{SEC:K}

 Mg does not show significant star-to-star dispersion in all but a handful of GCs (\S~\ref{SEC:CN}). In only two clusters  (namely NGC~2419 and, to a lesser extent, NGC~2808), low Mg abundances are also correlated with extreme K enhancements \citep[e.g.][]{Mucciarelli:12N2419,Carretta:15N2808}, whereas star-to-star scatter in K are not generally observed for the bulk of GCs \citep[][]{Takeda:09}. The K overabundance of Mg-poor stars can be produced, under some assumptions, by AGBs \citep[e.g.][]{Ventura:12}. However, both Na and Al are destroyed at the typical temperatures at which K is produced, e.g. Na and K are anti-correlated in stellar ejecta \citep[][]{Prantzos:17}. Thus, the simultaneous Na and K enrichment seen in NGC~2419 and NGC~2808 cannot be explained if the observed Na and K inhomogeneities are produced by the same stellar source.  As NGC~2808 and NGC~2419 are unusual in terms of the K-abundance patterns it is not clear if this is a promising window into the MP phenomenon, or instead pathological cases that confuse the issue.

\subsubsection{Multiple Populations in Extragalactic Environments}

MPs  have been also found outside our Galaxy. Star-to-star abundance variations in N, Mg, Na, and Al were  reported in extragalactic GCs by \citet{Mucciarelli:09}, who studied three ancient GCs in the LMC \citep[see also][for earlier studies]{Letarte:06,Johnson:06}. They found that these three clusters followed the same Na-O and Al-Mg anti-correlation trends as seen in Galactic GCs.  \citet{Hollyhead:17} measured the N and C-abundances of stars in the $\sim8$~Gyr SMC cluster, Lindsay 1 based on low resolution spectroscopy of cluster members.  
Using HST imaging in filters that are sensitive to C, N and O variations (see Fig.~\ref{fig:SPECTRA}),  \citet{Larsen:15Fornax}  determined the presence of MPs in four GCs in the Fornax dwarf spheroidal galaxy; they have also been detected in three 6 - 8 Gyr clusters in the SMC \citep{FlorianSMC}; as well as in the only {\em classical} GC in the SMC 
\citep[][]{Dalessandro121,Florian:121}.

\begin{marginnote}[]
\entry{SMC}{Small Magellanic Cloud}
\entry{LMC}{Large Magellanic Cloud}
\end{marginnote}

There are a number of GCs within the Milky Way that likely originate from accreted dwarf galaxies.  These include GCs associated with the Sagittarius dwarf galaxy, for example M54 (perhaps the nucleus of the galaxy, see ~\S~\ref{sec:omega}), Terzan 7 \& 8, Pal~12, \& Arp~2.  M54 certainly shows MPs \citep{Carretta:10M54}, while the situation is less clear for Terzan~7 and 8 and Pal~12 due to the small samples of stars observed in each \citep[e.g.][]{Cohen:04}.
In addition to resolved star studies, integrated light studies have also found strong evidence for MPs to be present in extragalactic clusters by looking for GCs that are strongly enriched in N or Na.  These include many ancient GCs in M31 \citep{Schiavon:13,Colucci:14,Sakari:15} and the lone GC associated with the WLM dwarf galaxy \citep{Larsen:14WLM}.

There have also also been attempts to search for MPs in extragalactic environments through integrated light photometry in the UV.  If (large) He spreads are present within the clusters, an extreme HB may develop causing significantly more UV emission than if all stars have the nominal He abundance.  Such UV-excess has been observed in some massive extragalactic GCs in M87, M31 and M81 \citep[e.g.,][]{Sohn:2006,Mayya:13,Peacock:17}.

Based on these studies, along with those of Galactic GCs, it appears that one of the main properties of MPs is their near ubiquity in ancient and massive GCs \citep[c.f.,][]{Renzini:15}.  However, as will be discussed in \S~\ref{sec:age_mass} this near ubiquity does not appear to apply to the young and intermediate age ($\lesssim2$~Gyr) massive clusters in the LMC/SMC.

\subsection{Multiple Populations as Seen Through CMDs}\label{SEC:PHOTOMETRY}
The peculiar MP chemical composition can also be seen through accurate photometry \citep[e.g.][]{Hartwick:72}.
Imaging allows us to discriminate efficiently between 1P and 2P sub-populations through photometry in samples composed of many thousand of stars, while simultaneously covering a wider region in the sky (a result that is difficult to achieve with the most advanced spectroscopic facilities, even for nearby clusters). The relative number ratios between 1P/2P stars can be inferred and the radial distribution of the two groups can be investigated in detail by taking advantage of the large number statistics secured through photometry \citep[e.g.;][]{Lardo:11,Lee:17}.
Nonetheless, wide-field photometric observations covering the full extension of the clusters (i.e., out to the tidal radius) are available only for a subset of clusters \citep[][]{Dalessandro:14,Massari:16} even if a large amount of archival data are publicly available in the archives. 

HST offers very high precision and accuracy to effectively sort different sub-populations \citep[][]{Piotto:15UVsurvey,Soto:17,Milone:17}.  The {\em HST UV Legacy Survey of Galactic Globular Clusters} (PI G. Piotto) has had a major impact on the field, allowing for the exploration of MPs and the link with their host cluster in unprecedented precision. However, space-based observations have only a limited spatial coverage\footnote{Moreover, different regions of the clusters are included in the HST FoV, depending on the specific properties of the cluster itself, i.e. core/half-light radii and heliocentric distance.}. 
The less dense outer parts of clusters (where the two-body relaxation timescale is longer and mixing less efficient) can retain imprints of different initial configurations of MPs as differences in their relative spatial distributions or kinematics, hence their study allows us to gain crucial insights on the dynamics in play at the formation of the different sub-populations.

\subsubsection{Causes for the Complex CMDs and Filter Dependence}
Splits or spreads in cluster CMDs have been used to identify MPs and constrain their properties. The cause of these splits
depends on the colour (or colour combination) used to image clusters and on the specific evolutionary stage considered. Briefly, filters encompassing  wavelengths shorter than $\sim$ 4000\AA~are very sensitive to individual variations of  C, N, O in the outer layers of stars with cooler atmospheres. Conversely, star-to-star variations in He (as well as the CNO sum) impact primarily the stellar structure. As such, they affect mainly optical bands although they have some influence on the UV.

\citet{Salaris:06} firstly considered the effect of He and light element variations on photometry. They conclude that in the Johnson-Cousins B,V, and I filters only an extreme helium enhancement (Y$\geq$ 0.35) leads to an appreciable colour change of stars with 2P composition as compared as a standard 1P stars. A prominent splitting of the MS and the MSTO is produced by relatively large He enhancements, while colour variations due to He variations are less pronounced in the RGB in optical colours. The CNONa anti-correlations do not affect the evolutionary properties of stars, hence the position of stellar models in the theoretical H-R diagram, when the C+N+O sum is kept constant \citep[][]{Sbordone:11}. On the other hand, the observed splitting of the SGB into a brighter and fainter sequences in some clusters in optical filters can be interpreted as the result of a change in the C+N+O sum \citep[][]{Cassisi:08,Piotto:12}. Moreover, 1P and 2P stars have also slightly different luminosity at the RGB bump and they occupy different regions on the HB when clusters are imaged with optical BVI filters \citep[e.g.; ][]{Bragaglia:10a}.

\begin{figure}
\includegraphics[width=0.74\columnwidth]{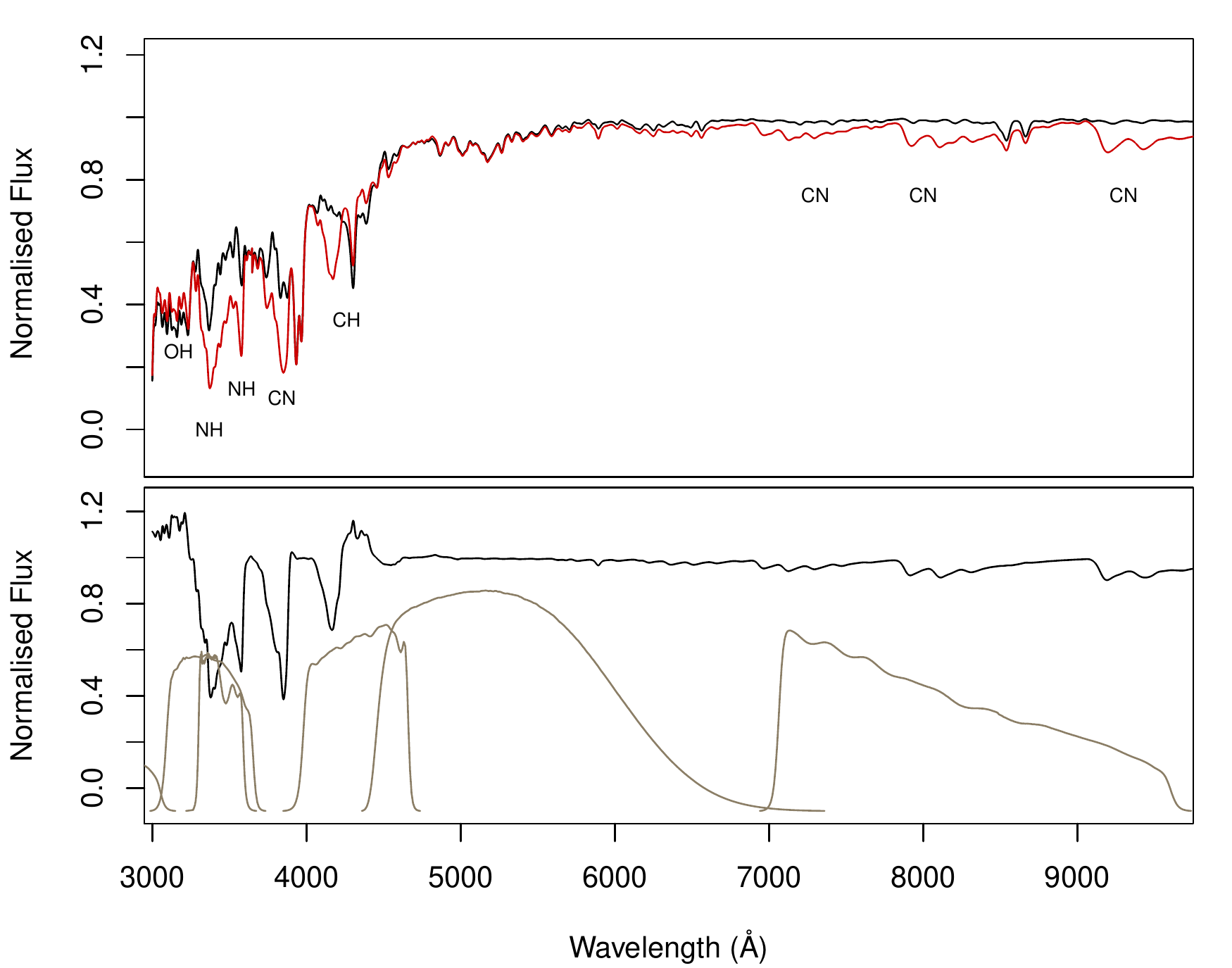}
\caption{Normalised synthetic spectra of RGB stars with 1P and 2P composition are plotted in the top panel.  A number of molecular absorption bands that vary significantly between the two spectra, are also labelled.
The flux ratio between the two spectra ratio is shown in the bottom panel, along with some WFPC3/UVIS filters used in photometric studies to pinpoint the presence and properties of MPs
(bottom panel, from left to right: F336W (U), F343N, F438W (B), F555W (V), F814W (I), where the we also list the approximate Cousins-Johnson filter equivalent in parenthesis).  After \citet{Sbordone:11}.}
\label{fig:SPECTRA}
\end{figure}

Larger colour spreads (from the MS up to the RGB, where the effect tends to be larger) are expected in CMDs including near ultraviolet filters, even while leaving the C+N+O sum unchanged \citep[see][for a comprehensive discussion]{Petrinferni:09}. C, N, and O individual variations are critical, while He enhancement works in the opposite direction of CNONa spreads.
This property appears to be shared by any filter encompassing the wavelength range 3000 $\leq$ $\lambda$ $\leq$~4000\AA, where most of the NH and CN absorptions are located.  In Fig.~\ref{fig:SPECTRA} we show synthetic spectra of RGB stars with typical 1P (black) and 2P (red) chemical abundances, and highlight molecular bands that differ between the spectra.  Additionally, in the bottom we show the flux ratio between the two spectra and the throughput curves of selected HST filters.  Due to these spectral differences the colour spread observed in specific colour combinations including near-UV filters has been shown to be very sensitive to light elements abundances \citep[e.g.][]{Marino:08M4}. Several combinations of colours have also been introduced to best disentangle the different subpopulations.
For example, \citet{Monelli:13} found that all of the 23 clusters in their sample analysed with ground-based photometry show broadened or multimodal RGBs in the C$_{\rm U,B,I}$ = (U -- B) -- (B -- I) vs. V CMDs, where the different branches of the RGBs are tightly linked to their light element content (see Figs.~\ref{fig:CUBI} \& \ref{fig:SPECTRA}).   \citet{Florian:121,FlorianSMC} imaged a number of clusters in the LMC in the colour index C$_{{\rm F336W,F438W,F343N}}$ = (F336W -- F438W) -- (F438W -- F343N) to pinpoint the presence of MPs with different C and N abundances, finding evidence for MPs for all observed clusters older than $\sim$ 6 Gyr \citep[see also][]{Hollyhead:17}.

In Fig.~\ref{fig:CUBI} we show an example case of NGC~6752.  In the left panel we show the C$_{\rm U,B,I}$ CMD showing the split/spread RGB of the cluster in this filter combination.  Additionally, we show the position of stars on the RGB, labelled in terms of their chemical abundances (right panels).  Hence, the position of a star in CMDs, in specific filter combinations, can be used to trace the chemical composition of the stars.

\citet{Milone:17} used a similar colour index to constrain the presence and properties of MPs in 57 Galactic old clusters
using the large database of data coming from the HST Large Program {\em The HST Legacy Survey of Galactic Globular Clusters: Shedding UV Light on Their Populations and Formation} \citep[see][]{Piotto:15UVsurvey,Soto:17}. UV observations  taken in the F275W, F336W and F438W filters further complement optical HST observations from the  {\em ACS Survey of Galactic Globular Cluster} \citep[e.g.][]{Sarajedini:07} with WFC3/UVIS images. The defined C$_{{\rm F275W,F336W,F435W}}$ = (F275W -- F336W) -- (F336W -- F435W)  colour combination allows one to clearly identify photometric splits/spreads caused by variations in individual elements, namely C, N, and O (see Fig.~\ref{fig:SPECTRA}). Also, the combination of UV CMDs with optical photometry allows He enhancement ($\Delta$Y) of the different subpopulations to be seen.

A pseudo colour-colour diagram (or chromosome map; see Fig~\ref{fig:uv_hst_legacy}) has also been  introduced to 
identify different subpopulation from the HST UV survey photometry by highlighting subtle chemical differences (in light elements and He)
between them \citep[e.g.][]{Milone2808He}. Briefly, two fiducial lines are drawn to fit at the blue and red envelope of the RGB sequence in the 
F814W vs.  C$_{{\rm F275W,F336W,F435W}}$  and F814W vs. (F275W-F814W) CMDs.
The red and blue fiducial lines are then used to verticalise the RGB sequence in a way that they translate into vertical
lines.  A pseudo colour-colour plot can then be made of the position of each RGB star in the verticalised colours, $\Delta ^{N} _{C\rm {F275W,F336W,F438W}}$ and $\Delta ^{N}_{{\rm F275W,F814W}}$. An example of such a diagram can be seen in Fig~\ref{fig:uv_hst_legacy} for NGC~2808 stars, which reveals the presence of at least six sub-populations with distinct chemistry.

With such diagrams, \citet{Milone:17} were able to efficiently distinguish the 1P and 2P populations for most clusters, although some clusters did display a continuous distribution (see Fig.~\ref{fig:uv_hst_legacy} for the division).  These distinctions were confirmed through comparison with the results of ground based spectroscopic studies, i.e. 1P stars identified photometrically corresponded to stars with the field abundance patterns of Na and O.  

\begin{marginnote}[]
\entry{YMC}{Young Massive Cluster - a.k.a. young GC}
\end{marginnote}

With the precision of HST photometry, relatively tight constraints can be placed on any age difference between the populations.  Using the UV-Legacy survey data, \citet{Nardiello:15} selected stars from the 1P and 2P populations based on UV images in the Galactic GC NGC~6352.  The authors then estimated the age of each population independently, using optical CMDs (V-I vs. I) centred on the MSTO of each population.  The optical colours are not strongly affected by MPs (although He variations can affect optical colours as well as non-constant C+N+O sums) hence any differences would be attributed primarily to age differences (if He variations are taken into account, which the authors did).  In this case, the age difference was found to be $10 \pm 110$~Myr.  When all sources of uncertainties are included (including the [$\alpha$/Fe] ratio), the authors find that the two populations are coeval with an upper limit of $300$~Myr between them.  This is consistent with a similar upper age limit found by \citep[][]{Marino:12} for M22.  Tighter age constraints can be gotten from younger clusters that show MPs (YMC; see \S~\ref{sec:ymcs}).

\subsubsection{A spread amongst 1P stars?}

An unexpected result of the \citet{Milone:17} study was that the 1P population displayed a significant spread in some clusters (although no spread was seen in Na and O for these stars) while being quite compact in other clusters.  
 Based on the data provided in \citet{Milone:17}, it appears that $\sim70$\% of the GCs in that sample display a significant spread in their 1P stars.  While this appears to be common, many clusters do not show an extended 1P, and it not clear at present what (if any) cluster property controls the spread in the 1P stars. 

\begin{figure}[!tb]
    \centering
    \begin{minipage}{.5\textwidth}
        \centering
       \includegraphics[width=1\linewidth]{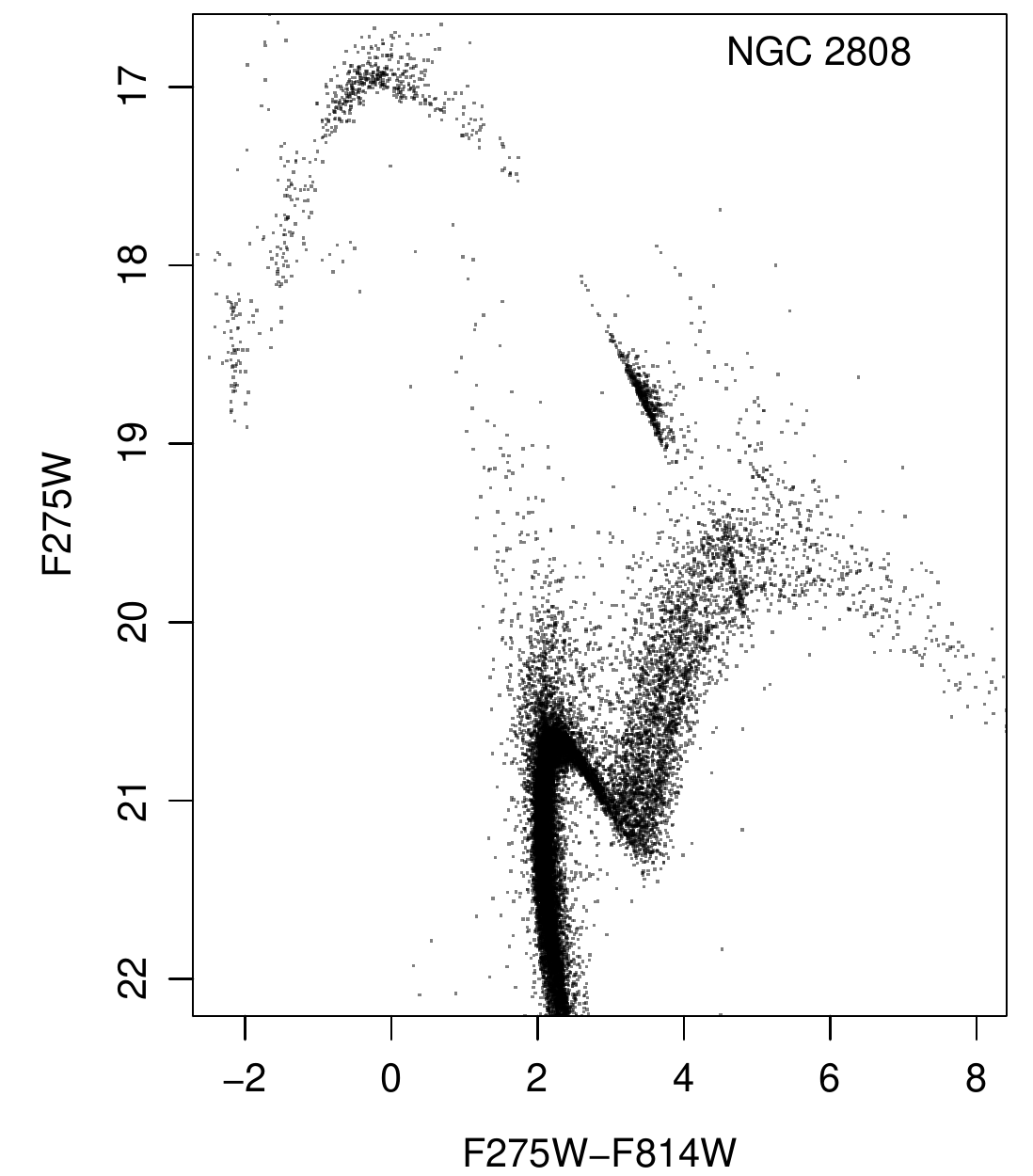}
    \end{minipage}%
    \begin{minipage}{0.5\textwidth}
        \centering
        \includegraphics[width=1\linewidth]{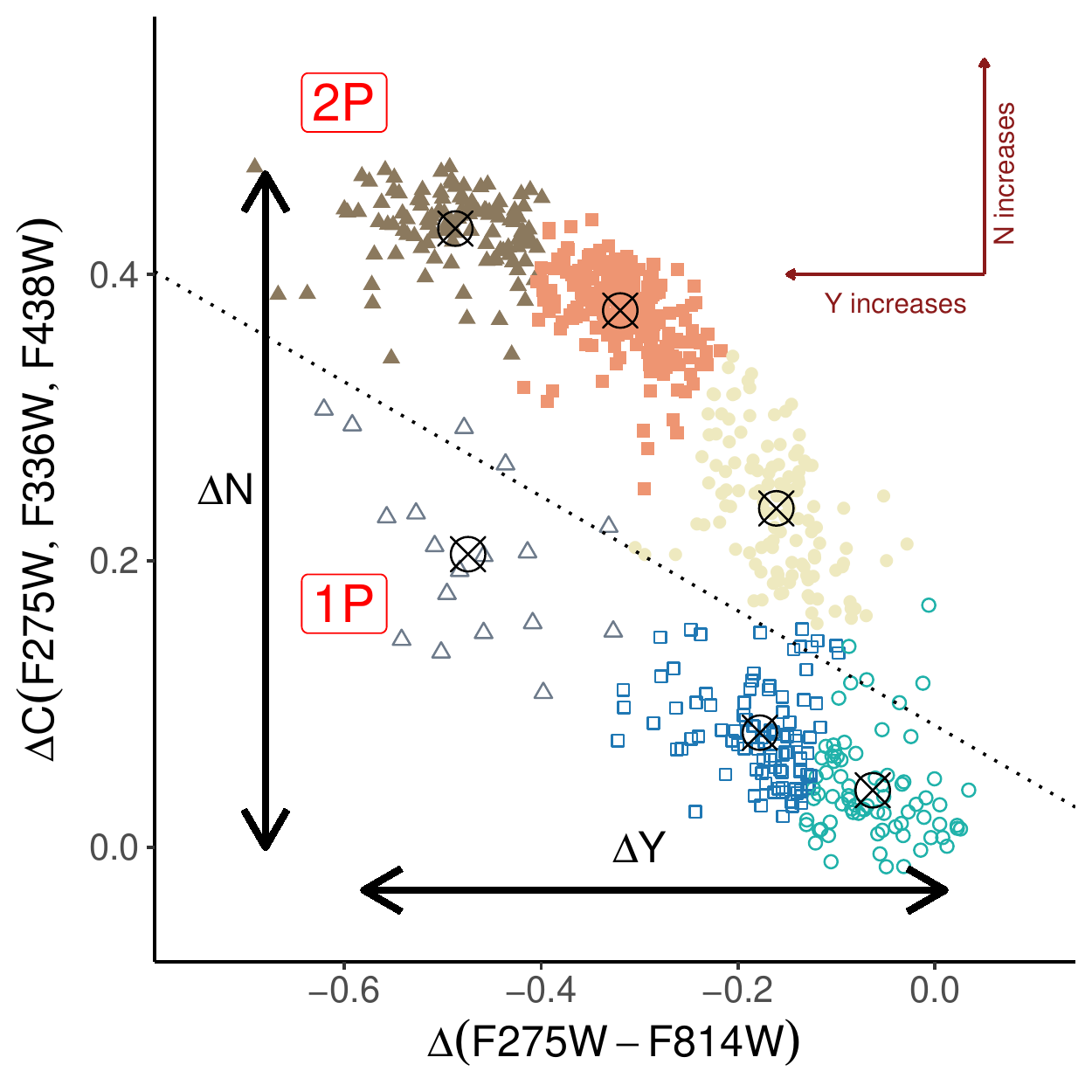}
    \end{minipage}
\caption{{\bf Left panel:} An HST UV-optical CMD of the central regions of NGC~2808 \citep[data are from][]{Piotto:15UVsurvey}.  Note the distinct multiple RGBs and the highly structured HB.  This complexity is due to light element abundance variations (He, C, N and O) between cluster stars. {\bf Right panel:} A ``chromosome map'' of NGC~2808 (after \citealt{Milone:17}) for RGBs (i.e. relative positions of the stars on the RGB in different filter combinations that are sensitive to different abundance variations) where at least six distinct populations can be inferred.  Here the x-axis is mainly sensitive to variations in He while the y-axis is dominated by variations in N (at C, O to a lesser extent). Based on the definition of \citet{Milone:17}, stars above the dashed line are considered 2P, while stars below the same line are 1P. Note that both the 1P and 2P  consist of 3 extended sub-populations.
}
\label{fig:uv_hst_legacy}
\end{figure}

Preliminary computations (Lardo et al., in preparation) reveal that for intermediate- and low- metallicities the $\Delta C(\rm {F275W,F336W,F438W)}$ colour spread essentially traces N (e.g., stars are sorted in order of increasing N abundance from bottom to top in the chromosome map of Fig.~\ref{fig:uv_hst_legacy}). Conversely, the $\Delta ({\rm F275W,F814W})$ colour spread is sensitive to He enhancement of the different subpopulations (e.g. in order of increasing He content, from right to left; see right-hand panel of Fig.~\ref{fig:uv_hst_legacy}).  The spread in 1P stars is seen predominantly in the $F275W-F814W$ colour (UV - I) suggesting that He variations are present within the 1P, which would be very surprising given the lack of Na, N or O variations within this population.  
This in turn suggests that some stars with little or no N-spread, show significant enhancement in their He values, which is in conflict with basic nucleosynthesis. Hence something else, other than the recycled by-products of stellar nucleosynthesis, has caused the He variations within 1P stars.  This appears to be a particularly promising avenue for future study.

\subsection{Are there single population GCs?}
Nearly all GCs analysed at high-resolution, with exception of Ter~7, Pal~12, Pal~3, and Rup~106, show the Na and O variations. Ter~7 and Pal~12 are low mass members of the Sagittarius and 
high-resolution abundances exist only for a handful of cluster members \citep[$\leq$ 5 stars;][]{Cohen:04, grazina:04,Sbordone:07}. The same holds for Rup~106, a slightly more massive (5 $\times$10$^{4}\msun$) cluster with a probable extragalactic origin \citep[9 stars;][]{Villanova:13Rup106} and Pal~3, a distant GC in the outer halo, where the available data (2 stars) can neither confirm nor refute the presence of a Na-O 
anti-correlation \citep{Koch:09}. Increasing the sample of stars studied in these low mass clusters is essential to determine if there is a lower GC mass limit where MPs are present \citep[e.g.,][]{Dalessandro:14}.  In this respect searching for MPs through photometric methods can be problematic in these clusters as the low number of RGB stars often makes it difficult to identify MPs there, unless the populations are well separated (i.e., have large N or He variations).

In this respect, the case of the SMC old cluster NGC~121 studied by \citet{Dalessandro121}, is quite illustrative.
The authors derived Na and O for five RGBs and found no intrinsic scatter in both elements.
However, they detected two RGB sequences in their UV images, meaning that MPs are present. 2P stars were missed in their spectroscopic sample as it was biased (as most spectroscopic samples are) to the outer regions of the cluster, where the fraction of 2P stars is often lower in than in the central regions.

Two other old GCs have been claimed not to host MPs based on either ground based photometry or low resolution spectroscopy, E~3 \citep[][]{Salinas:15} and IC~4499 \citep[][]{Walker:11}, although followup HST photometry has detected MPs in IC~4499 (Dalessandro et al. in prep.).  Additional high-resolution studies designed to measure the abundance of the relevant light elements (e.g. Na, O, etc) for a representative number of stars in such clusters are needed to draw firm conclusions on the presence of MPs.

As will be discussed in \S~\ref{sec:age_mass}, a number of high mass ($\sim10^5$~\msun) clusters younger than $\sim2$~Gyr have been studied, and so far none have been found to host MPs (e.g., Mucciarelli et al.~2008; 2014; Martocchia et al.~2017).

\begin{figure}[!htb]
    \centering
        \centering
        \includegraphics[width=0.82\linewidth]{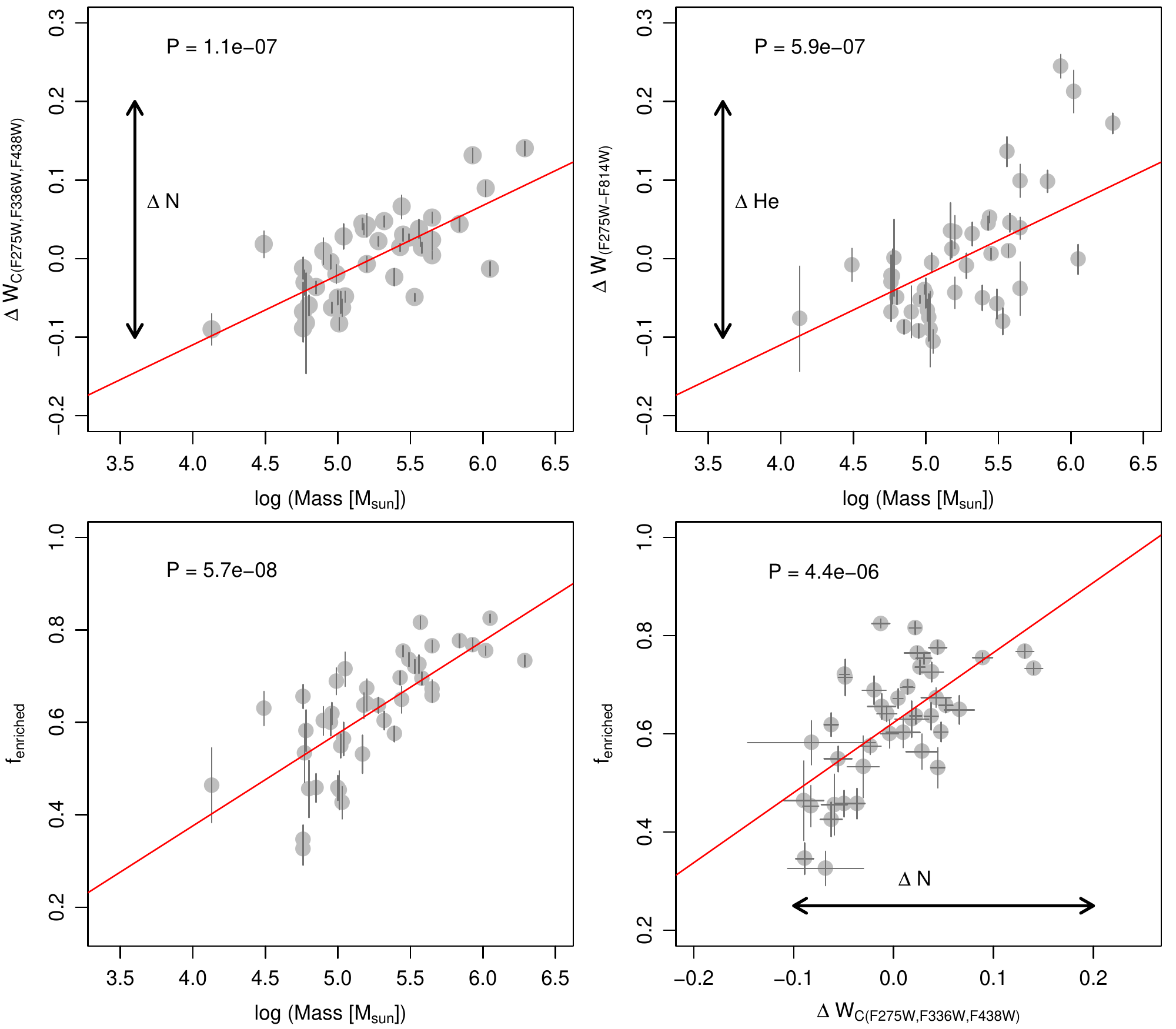}

 \caption{Based on results from the HST UV Legacy Survey we show a summary of how MP properties vary with the present day GC mass \citep[after][]{Milone:17}.  $\Delta W_{\rm F275W,F336W,F438W}$  
 and $\Delta W_{\rm F275W - F814W}$ are the widths of the RGB in the two colours (or colour combinations) corrected for the effect of metallicity, in a first approximation a measure of the amount of N- and He-enrichment (respectively) present in the cluster (i.e. the difference between the most enriched stars and the primordial stars). f$_{\rm{enriched}}$ is the fraction of 2P stars relative to the total number of stars, as measured on the RGB.   In the {\bf bottom panels} we show  f$_{\rm{enriched}}$ vs. cluster mass and $\Delta W_{\rm F275W,F336W,F438W}$. The solid (red) lines in each panel gives the best linear fit to the data, and the probability of no correlation between the points (P) is shown in each panel.  Self-enrichment scenarios (for standard nucleosynthetic stellar sources) all predict an anti-correlation between f$_{\rm{enriched}}$ and $\Delta W_{\rm F275W,F336W,F438W}$, opposite to the observed trend.  All data are from \citet[][]{Milone:17}.}
\label{fig:correlations_with_mass}
   
\end{figure}

\subsection{Global properties and correlations}
\subsubsection{Spatial Distributions, Dynamics, and Binary Properties of the Different Populations}
\label{sec:spatial_distributions} 

In many cases different stellar sub-populations seem to not share the same radial distribution. Across a range of cluster-centric distance, most studies have found that 2P stars are systematically more concentrated in the innermost region than 1P stars \citep[e.g.][]{Lardo:11,Simioni:16}. Only a few exceptions to this general trend have been reported, with stars with primordial composition being more centrally concentrated than 2P giants \citep[][]{Larsen:15m15,Vanderbeke:15,Lim:16} or 1P and 2P stars having the same radial distribution \citep[e.g.][]{Dalessandro:14,Miholics:15}.
Hints that 2P stars have lower velocity dispersion  \citep[e.g.][]{Bellazzini:12,Kucinskas:14} and more radially anisotropic velocity distribution \citep[][]{Richer:13,Bellini:15} have also been reported. The binary properties of 1P and 2P stars may also be different, with 2P stars showing a lower binary fraction \citep[][]{Dorazi:10,Lucatello:15}.

\subsubsection{Observed Population Ratios}

While there are radial trends in the 2P/1P ratios, in most cases large samples of stars are required to demonstrate this statistically.  Overall, 2P stars make up the majority of stars in most GCs, although the fraction of 2P stars is seen to be a strong function of cluster mass, with more massive clusters having larger fractions of 2P stars (e.g., Milone et al.~2017 - see Fig.~\ref{fig:correlations_with_mass}).  \citet{BastianLardo:15}, using mainly spectroscopic results from the literature which are biased towards the outer regions of clusters, did not find any trends between the enriched fractions (\fenrich $=N_{\rm 2P}/N_{\rm tot}$) and metallicity or galactocentric distance\footnote{They also did not find any correlations between \fenrich\ and cluster mass, but found an average value of $\fenrich=0.68$ which agrees well the with the average from HST photometry, although why they did not find a trend with mass is not entirely clear.}.  This has been confirmed with HST photometry \citep{Milone:17}.  Hence, the MP phenomenon is not directly linked to the environment in which the cluster forms (e.g., within dwarf galaxies or the bulge of the Galaxy).

The trend between population ratios and mass is a key constraint on scenarios for the origin of MPs which will be discussed in \S~\ref{sec:trends_cluster_properties}.

\subsection{The Role of Cluster Age and Mass}
\label{sec:age_mass}
\begin{figure}[!tb]
    \centering
    \begin{minipage}{.7\textwidth}
        \centering
       \includegraphics[width=1\linewidth]{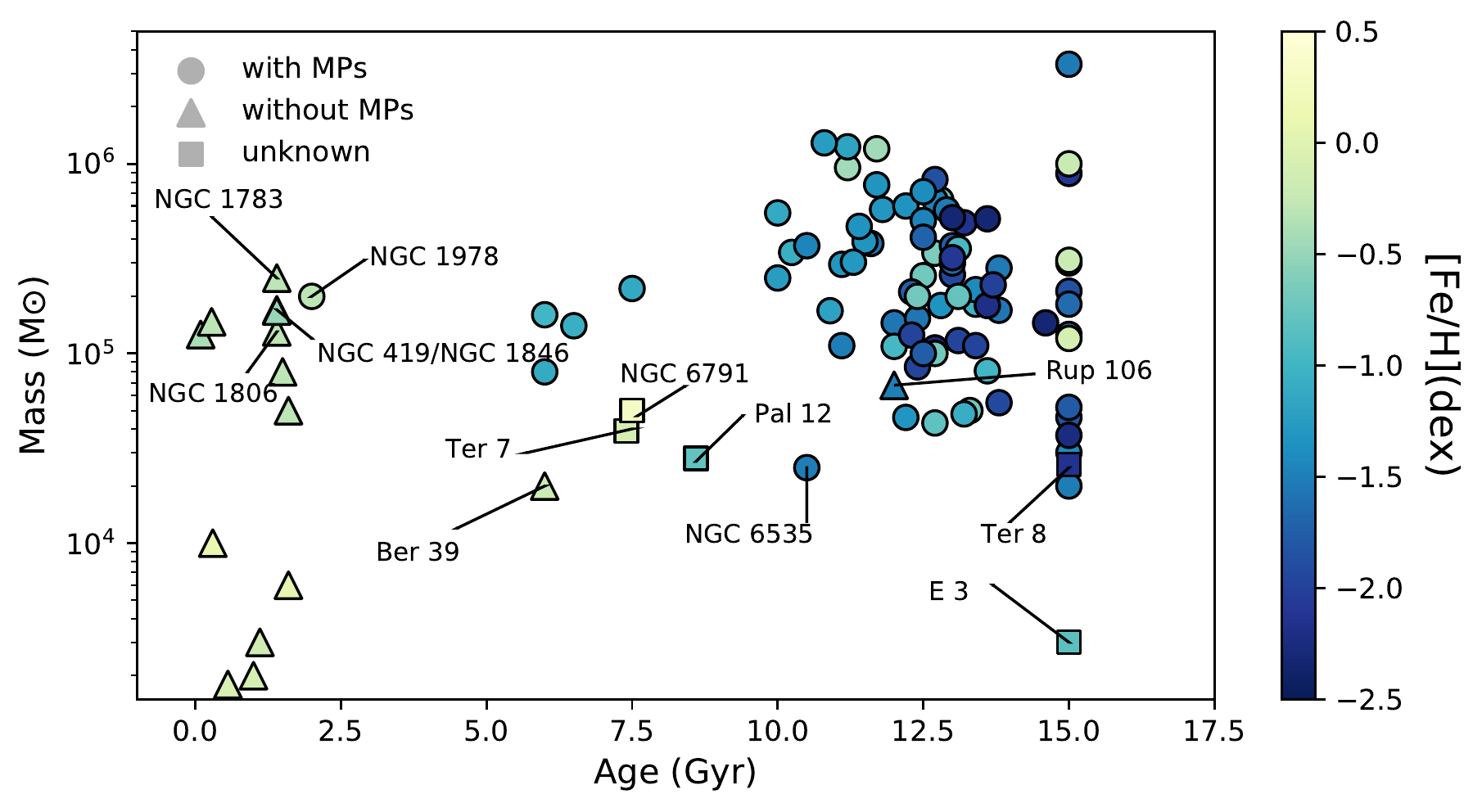}
    \end{minipage}
    \begin{minipage}{0.7\textwidth}
        \centering
        \includegraphics[width=1\linewidth]{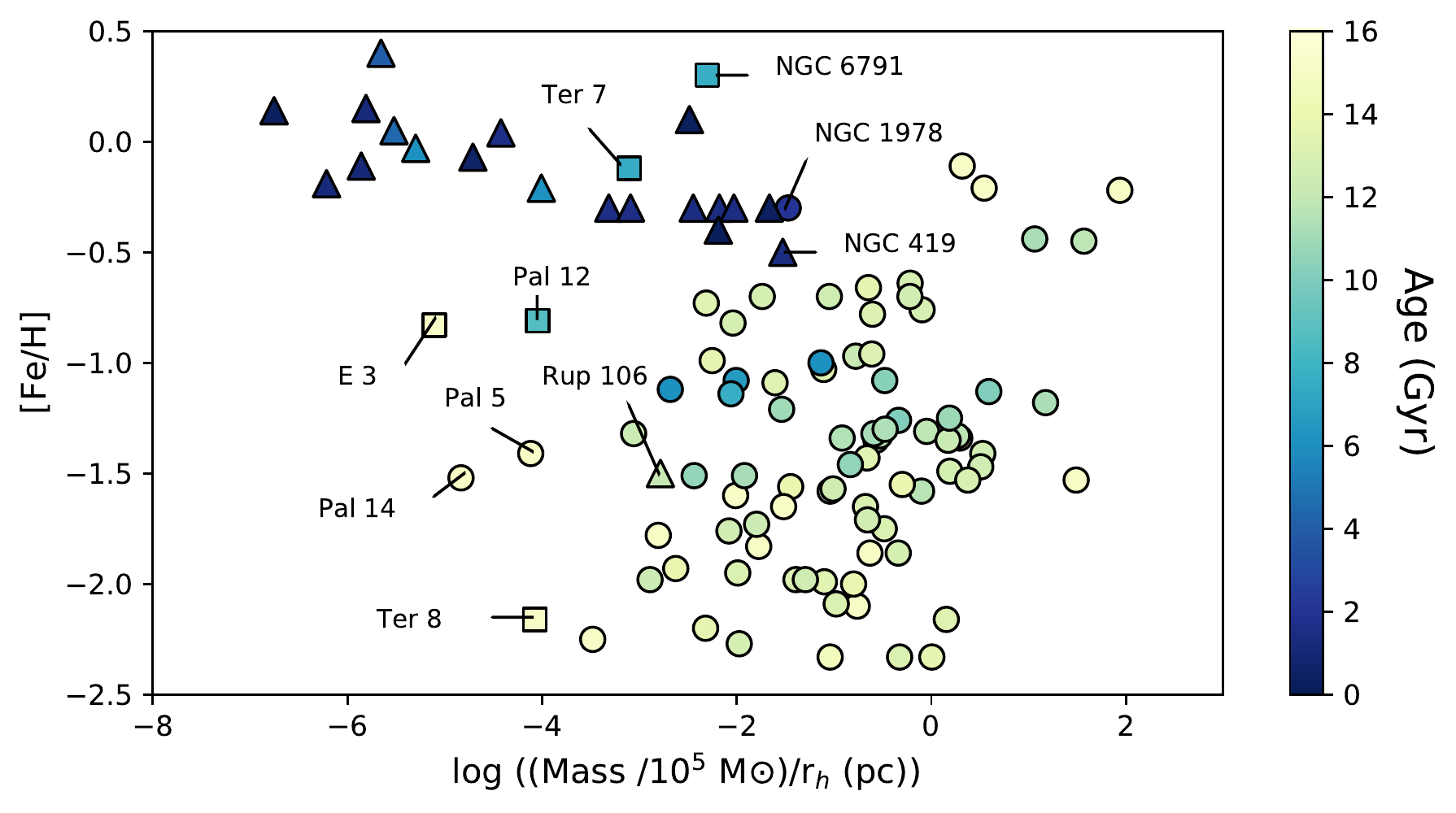}
    \end{minipage}
\caption{A summary of results from the literature on whether MPs are present within clusters.  Circles denote clusters where MPs have been unambiguously detected, triangles show where they have not been detected (with large enough samples to suggest a true absence) and squares show ambiguous cases (mainly due to small samples or potentially large observational uncertainties).  Some particularly interesting cases are labelled. An age of 15 Gy has been assigned to clusters for which no age determination has been found in literature.  Whether or not a cluster hosts MPs or not depends on its mass (or density) as well as its age.  The data  come from the compilation of \citet{Krause:16} with additional points added from more recent works discussed in this review.}
\label{fig:MASSAGE}
\end{figure}

It is still not clear precisely which properties of the clusters determine whether MPs will be present within the cluster.  However, with the release of large and homogeneous surveys we can begin searching for correlations between cluster properties (e.g. age, mass, location) and the presence/absence of MPs as well as their extent, in order to glean clues as to the mechanisms responsible for MPs.  In Fig.~\ref{fig:MASSAGE} we show a collection of clusters from the literature where MPs have been searched for, in the age-mass plane and the [Fe/H]-concentration (mass/R$_{\rm h}$) plane.

\subsubsection{Cluster Mass}

As it became apparent that (nearly) all of the ancient GCs host MPs and that (so far) none of the open clusters do, it was suggested that cluster mass may play a key role \citep[e.g.][]{Carretta:10GLOB}.  The general argument is that if clusters host a deep enough gravitational potential well, they may be able to retain the stellar ejecta of a first generation of stars and form a second generation with that enriched material.  This is generally based on an escape velocity argument although often overlooks the role of energetic stellar sources, like high/low mass x-ray binaries or ionising white dwarfs \citep[e.g.][]{Dercole:08,Krause:12,McDonald:15}.

Cluster mass does appear to be an important parameter for GCs, playing a role in determining whether MPs are present, but also in the properties (i.e. how severe the abundance variations are) of the MPs.  The first hints for this came from \citet{Carretta:10GLOB} who used their large sample of stars in 19 GCs to search for correlations between the extent of the Na-O anti-correlation (as measured through the interquartile distribution) and various cluster properties.  The strongest relation found was with cluster mass, with higher mass clusters showing larger Na-O abundance spreads.  This is difficult to reconcile with standard stellar evolution, as the stellar ejecta released into the cluster should not depend on cluster properties.  For models that invoke dilution, this would require that lower mass clusters undergo more dilution (whereas lower mass clusters would be expected to accrete less gas from their surroundings) or that higher mass clusters retained a larger fraction of the processed material (although models already adopt that all clusters retain 100\% of the processed material).  Since models already assume that GCs retain 100\% of the material processed through the enriching source (e.g., FRMSs, AGBs, IBs, etc) this will further exacerbate the mass budget problem (see \S~\ref{sec:mass_budget}).
 
Similarly, \citet{Milone:15M62} found that the He spread ($\Delta Y$) within Galactic GCs is much larger in higher mass clusters.  Although this was only based on nine GCs, it will be directly tested with a much larger sample from the UV Legacy Survey of GCs \citep{Piotto:15UVsurvey}.  In Fig.~\ref{fig:correlations_with_mass} we show the results from \citet{Milone:17} for the width of the RGB in the $(F275W-F814W)$ CMD (corrected for metallicity effects) which is a proxy for He spread (Lardo et al. in prep.).  This confirms and extends the trend reported by \citet{Milone:15M62}.
One of the major results from the UV GC Legacy Survey has been the discovery of a strong correlation between cluster mass and the fraction of enriched stars (\fenrich) within the cluster \citep{Milone:17}. Here, \fenrich\ is found in the $\Delta_{\rm{F275W,F814W}}$ vs. $\Delta_{\rm{C F275W,F336W,F438W}}$ colour-colour plot (see Fig.~\ref{fig:uv_hst_legacy}).  The authors note that in some cases the 1P population appears to be made up of multiple groups, hence \fenrich\ may be a lower limit.  In Fig.~\ref{fig:correlations_with_mass} we show some of the main results from \citet{Milone:17}, namely how N-spread ($\Delta_{\rm{C F275W,F336W,F438W}}$), He-spread ($\Delta_{\rm{F275W,F814W}}$), and f$_{\rm enriched}$ vary as a function of mass (after removing the trends with metallicity).

High mass clusters (e.g., NGC~2808, 47~Tuc with M$_{\rm cluster} \sim10^6$~\msun) have \fenrich$\approx 0.8$, while clusters with masses near $10^5$~\msun\ have \fenrich$\sim0.4-0.5$.  Note that the enriched population still makes up a substantial fraction of the stars even in low mass clusters.  It is not just the fraction of enriched stars that varies with cluster mass, it is also the extent of the enrichment as well (i.e. larger abundance spreads in higher mass clusters.  This is in agreement with earlier work based on spectroscopic samples \citep[][]{Carretta:10GLOB}.   The implications of these results will be discussed in \S~\ref{sec:trends_cluster_properties}.

There has also been studies focused on old open clusters, which typically have masses much lower than GCs, e.g., Ber~39 \citep{Bragaglia:12}.  To date, MPs have not been found in open clusters with masses as high as $2\times10^4$~\msun\ and ages as old as $\sim$ 6 - 9 ~Gyr.  Comparison of clusters with ages of 6-8~Gyr clusters in the SMC with masses of $\sim10^5$~\msun\ \citep{Hollyhead:17,FlorianSMC} with their lower-mass open clusters counterparts (e.g., Berkeley 39) hint that mass may indeed play a role (see Fig.~\ref{fig:MASSAGE}). The SMC clusters do host MPs, while open clusters do not.  Although, of course, the formation environment may also have been different.

Recent studies have also targeted low mass ancient GCs, such as NGC~6362 ($M\sim5\times10^4$~\msun, \citealt{Dalessandro:14}) or E3 ($1.4\times10^4$~\msun\ - \citealt{Salinas:15}), with mixed results.  NGC~6362 does host MPs, but based on its orbit and observed stellar mass function, it is likely that it has lost a significant amount of mass during its evolution \citep[e.g.,][]{Kruijssen09}.  E3, on the other hand, does not appear to host MPs, based on CN low resolution spectra.   The very extended (R$_{\rm h}\sim25$~pc) outer halo cluster Palomar 14 with a mass of only $\sim10^4$~\msun\ does appear to host MPs \citep{Caliscan:12}.  The current record holder for the lowest current stellar mass cluster that still hosts MPs is NGC~6535 with few$\times10^3$~\msun\ (\citealp{Milone:17}; Carretta et al. 2018).

A summary of the role of mass (and concentration) in whether a cluster hosts MPs or not is shown in Fig.~\ref{fig:MASSAGE}.  There is overlap between ancient GCs that do host MPs and younger clusters that do not. However, the data are consistent with a lower initial mass limit of $\sim10^5$ where MPs can develop (at least for clusters older than $\sim2$~Gyr, see next section).

\subsubsection{Cluster age and metallicty} 

As discussed above, nearly all of the ancient GCs that have been studied in the necessary detail host MPs. However, there are stellar clusters that formed after the peak epoch of GC formation ($z=2-5$), continuing to form up to the present day, that have masses and densities comparable to, or even significantly above, the ancient GCs.  Hence, an obvious question is whether these clusters also host MPs, and if so, can they be used to test the formation scenarios that have been put forward (see \S~\ref{sec:models}).

There have been a number of studies to search for MPs in massive clusters with ages $<8$~Gyr (see \citealt{Krause:16} for a recent review).   With only a handful of exceptions (discussed above) it appears that all massive clusters older than $\sim6$~Gyr host MPs \citep{Hollyhead:17,FlorianSMC} while all clusters younger than $\sim$2~Gyr do not \citep[e.g.,][]{Mucciarelli:08,Mucciarelli:14LMC,Martocchia:17N419}, even with mass being held constant (at $\sim10^5$~\msun; see Figure~\ref{fig:MASSAGE}).  

The $\sim6$~Gyr clusters, NGC~339, NGC~416 and Kron~3, all located in the SMC, show clear evidence for MPs \citep[][]{FlorianSMC}.  This age corresponds to a formation epoch of $z_{\rm form} = 0.65$, arguing against a cosmological origin of the phenomenon (i.e. special properties of the early universe that contributed to the formation of MPs).  Unexpectedly however, another massive cluster in the SMC, at an age of $\sim1.7$~Gyr, NGC~419, with a similar mass of $\sim2\times10^5$~\msun~does not host MPs, based on HST photometry \citep{Martocchia:17N419}. The youngest cluster found so far to host MPs is NGC~1978, at an age of $\sim2$~Gyr \citep{Martocchia:18a}, suggesting that MPs (at least on the RGB) develop in an extremely narrow age range (or alternatively stopped being able to form in the LMC/SMC) between $\sim1.7-2$~Gyr \footnote{We note, however, that in the $2-8$~Gyr clusters, MPs have only been identified through N-variations. High-resolution studies to also estimate Na and O in these stars would be a welcome contribution.}. This is shown in Fig.~\ref{fig:MASSAGE}, where clusters like NGC~1783 and NGC~1978 lie on opposite sides of this dividing line in age, although with nearly identical masses. However, there are also older clusters like Ber~39 (a Galactic OC ) that do not host MPs, suggesting that mass (and potentially formation environment) plays a strong role as well.

There have also been a number of studies that have searched directly for abundance spreads in young/intermediate age massive clusters, based on high-resolution spectroscopy \citep[e.g.,][]{Mucciarelli:08,Mucciarelli:14LMC} of individual stars.  No solid evidence for abundance spreads has been found so far for any cluster less than $\sim2$~Gyr.

A number of studies have attempted to search for abundance anomalies through integrated light spectral studies 
\citep[e.g.,][]{Colucci:12,CabreraZiri:16,Lardo:17Antennae}.  These are mainly focussed on finding high mean levels of elements that typically vary due to MPs, namely [Na/Fe] or [Al/Fe].  As with the resolved studies, to date there have not been clear indications for abundance spreads in the young or intermediate age clusters ($<2$~Gyr), although the ancient GCs do show the expected trends in integrated light.

Finally, Fig.~\ref{fig:MASSAGE} also shows the results from the literature on whether a cluster hosts MPs in [Fe/H] vs. concentration (mass/radius) space.  There is overlap in both [Fe/H] and concentration where clusters do/do not host MPs.  Systematic searches for MPs in diffuse GCs may lead to significant new insights.

\begin{summary}[Observational Summary of Multiple Populations]
\begin{enumerate}
\item MPs, as seen in light element abundance spreads (C, N, O, Na, Al, He and sometimes Mg), are nearly ubiquitous in old massive GCs, independent of their formation environment (formed within the Galaxy or elsewhere) or metallicity.  
\item  MPs can be defined through clear correlations and anti-correlations between light-elements.  The main ones being a Na-O anti-correlation, a N-C anti-correlation, a Na-N correlation, and N and Na being correlated with He.  In some clusters Li is correlated with O (and hence anti-correlated with Na), Li measurements are relatively scarce. 
\item In most clusters [Fe/H] is constant between the populations and the sum of C+N+O is also typically constant within the measurement uncertainties (although there are more clusters with C+N+O spreads than those with [Fe/H] spreads).
\item Observed abundance trends are qualitatively consistent with those expected from the yields of hot hydrogen burning (increase in He, N, Na, sometimes Al; decrease in C, O, sometimes Mg), however no nucleosynthetic source provides a quantitative match to the data simultaneously.  
\item It is the spreads in He, C, N, and O (mainly) that cause the complexity observed in high precision CMDs for the majority of clusters (i.e. not age spreads or Fe spreads).
\item The fraction of enriched stars (ranging from $40-90$\% in the ancient GCs), the extent of the anti-correlations, and the He spread within the clusters are all a strong function of the cluster mass, all increasing with increasing mass.  Hence, the cluster properties appear to play a strong role in the formation of MPs.  2P stars make up the majority of stars in most GCs today, meaning that a substantial amount of processed material is required to form them.  This leads to the "mass-budget problem" which will be discussed in \S~\ref{sec:mass_budget}.
\item  It appears that the abundance patterns are discrete, when high precision measurements are possible, with many clusters showing the presence of  $>3-4$ sub-populations.
\item The majority of clusters in the HST UV Legacy Survey ($\sim70\%$) show a spread in their 1P stars, in addition to the spread in the 2P stars.  Preliminary modelling suggests that this is mainly due to He variations in 1P stars that are not accompanied by variations in other light elements (e.g., N, Na, O).
\item  In most clusters studied to date the enriched population of stars is either more centrally concentrated than the primordial population or if the cluster is dynamical relaxed, the two populations share the same distribution.  However, in a handful of cases the situation is reversed, with the 1P stars more centrally concentrated than the 2P stars.
\item MPs have been detected in clusters as young as $\sim2$~Gyr, which corresponds to a formation redshift of $z=0.17$, well past the peak epoch of GC formation ($z=2-5$).  Surprisingly, MPs have not been found in massive clusters with ages less than $2$~Gyr.
\item MPs are found in the full range of GC metallicities, from [Fe/H]$\sim-2.5$ to near solar metallicity.

\end{enumerate}
\end{summary}

\section{Nucleosynthesis and Multiple Populations}\label{SEC:NUCLEARREACTION}

All elements whose abundances show considerable scatter in GC stars (i.e. C, N, O, Na, Mg and Al) may participate in hydrostatic hydrogen burning. As a consequence, the presence of  the C, N, O, Na, and Al anti-correlated ranges observed in GCs has been interpreted as the results of hydrogen-burning through the CNO-cycle and the NeNa- and MgAl-chains \citep[e.g.][]{Langer:93}.
In the CNO-cycle, H is converted into He, and the individual abundances of the C, N, and O are altered 
whereas their net sum remains constant (as required by observations, see~\S~\ref{SEC:CN}).
The CNO-cycle is activated at T $\sim$ 20 MK, while the NeNa chain requires temperatures around $\sim$ 40 MK. Na 
reaches its equilibrium value at $\sim$50 MK and decreases at higher temperature. At higher temperatures (T$\simeq$ 70 MK) Al can be produced by p-captures on Mg isotopes \citep[e.g.][]{Denisenkov:89,Prantzos:07}. 

Three stellar types have been proposed as candidate polluters, because they reach extreme temperatures within their interiors (see also \S~\ref{SEC:LI} and \S~\ref{sec:SCLUSTERS} for additional constraints from elements others than CNO, Na, Al, and Mg). The possible 2P processed material donors are: intermediate mass ($\sim$ 3-8 $\msun$) AGB stars experiencing HBB \citep[e.g.][]{Dantona:16}; massive stars $\geq$ 15 $\msun$ \citep[][]{Krause:13,deMink:09}\footnote{This happens in the cores of massive stars, so additional processes are necessary to bring the material to the surface.  In the case of single stars, rotational mixing has been suggested, the so-called FRMSs.  Interactions between massive stars in binary systems can also bring processed material to the surface.}, and VMS \citep[$\sim$ 10$^{4}$ $\msun$;][]{Denissenkov:14}. Scenarios where the mixed contributions by different polluters have also been proposed \citep[e.g.,][]{Sills:10,Bastian:13}.

\begin{figure}[!tb]
\includegraphics[width=0.95\columnwidth]{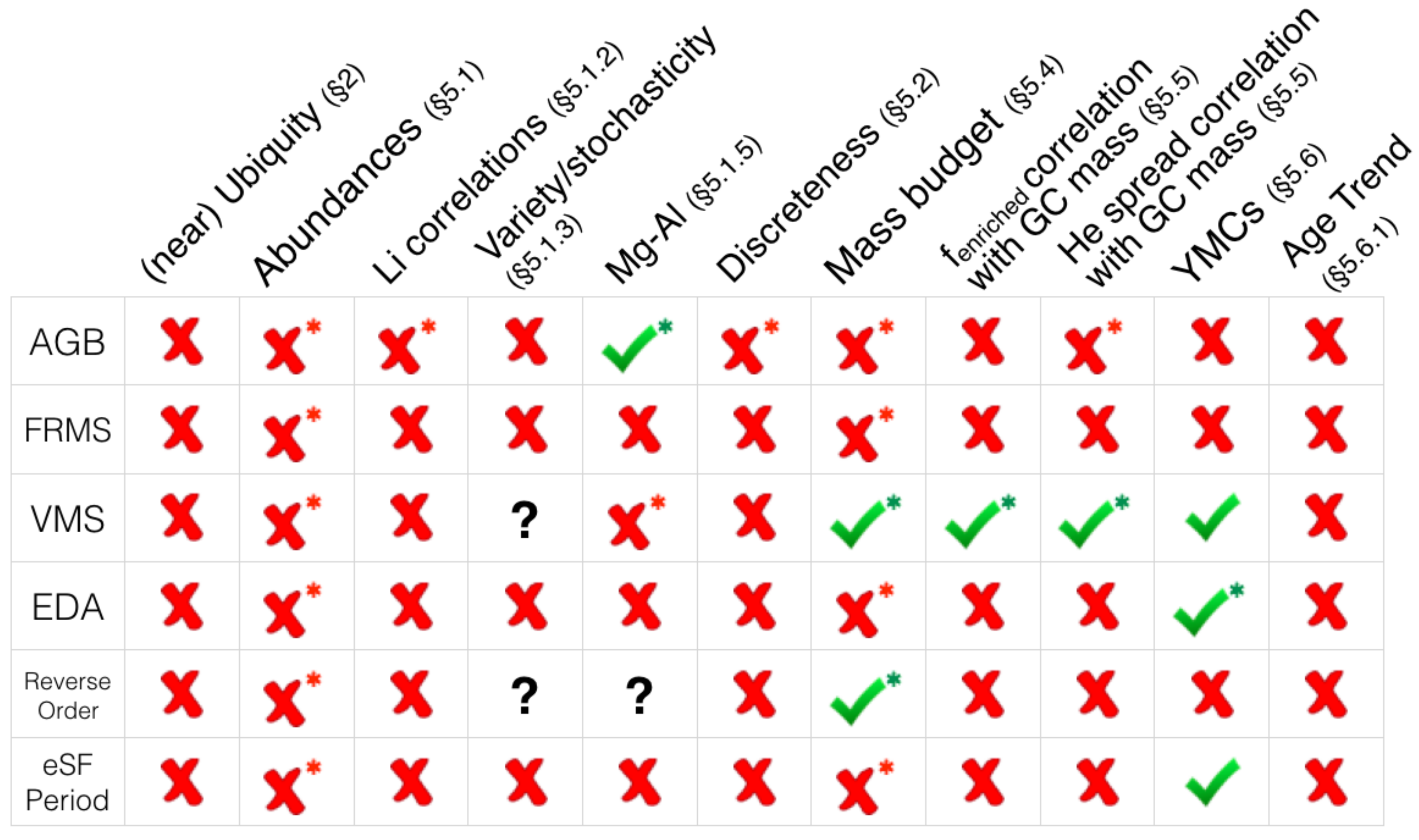}
\caption{A graphical summary of the comparison between predictions for the proposed models and observations (a.k.a. ``Truth Table").  A (red) cross shows a direct contradiction; a (red) cross with an asterisk shows a contradiction that may be avoided with relatively extreme fine tuning, or if the solution to that problem would violate another constraint; a (green) checkmark denotes where the prediction of a model is consistent with observations; a (green) checkmark with an asterisk indicates a situation where the model can be brought into agreement with observations with a (potentially) reasonable assumption (i.e. some degree of fine tuning is necessary); and finally a ``?" indicates where a model has not been developed enough to make a reliable prediction.  As can be seen, no model does particularly well when compared to observations.}
\label{fig:truth_table}
\end{figure}

As we discuss the characteristics of each of the proposed stellar sources as well as the scenarios developed around them, we will keep track of their successes and failures to reproduce key observations.  This will be done in Fig.~\ref{fig:truth_table}.  When a model matches an observation a green check will be used, while a green check with asterisk notes that the model may be consistent with observations under reasonable (assumptions).  Red crosses indicate when a model is in direct conflict with an observation and a red cross with asterisk shows where a model may match an observation but requires a high degree of fine tuning or by solving that problem it would violate another constraint.

\subsection{Potential Sources of the Enriched Material and Constraints from the Observed Variations}
 
Several observational constraints can naturally be reproduced within the proposed self-enrichment scenarios. Yet, a number of ad hoc assumptions must be made to explain other MP properties. For the sake of clearness, in what follows, we briefly introduce and discuss candidate stellar polluters for GC self-enrichment \citep[see also][]{Renzini:15,Charbonnel:16}.

\subsubsection{Massive Stars}
Massive ($\geq$ 15 $\msun$) MS stars reach the high temperatures required to manufacture the observed CNONaAl pattern very early in their MS evolution \citep[e.g.,][]{Maeder:06}. The fast rotation required by MP models allows for the transport of nuclides from the convective core to the radiative envelope, while losing mass through {\em (1)} a slow outflowing equatorial disc produced by a mechanical wind when the MS star rotates
close to critical velocity, and {\em (2)} a fast radiatively driven wind in the direction unhampered by the disc.  The enriched second generation stars a then predicted to form within this outflowing equatorial disc (i.e., a decretion disc).

  \begin{marginnote}[]
\entry{Decretion disc}{A disc made up of lost material around the equator of a rapidly rotating star}

\end{marginnote}

\begin{itemize}
\item The N-C and Na-O anti-correlated pattern is quickly established in massive star interiors, although the details of chemical enrichment depends on the adopted reaction rates. 
The FRMS are also able to process some Mg, which results
in a production of Al, at the expense of $^{24}$Mg. However,
this requires that the nuclear reaction rates for proton capture on
 $^{24}$Mg are increased by three orders of magnitude \citep[e.g.][]{Decressin:07a}. Using nominal reaction rates, 
FRMS would produce a positive Al-Mg correlation, which
 contradicts the observed anti-correlation.
Finally, the temperatures reached in massive star interiors are not high enough to build the Si-Mg anti-correlation observed in a subset of clusters (\S~\ref{SEC:CN}), nor variations in elements heavier than Al. 

\item Na and Al directly correlate with He, as observed (\S~\ref{SEC:HE}). However, the predicted He enhancement is significantly higher than the value allowed by observations \citep[see \S~\ref{SEC:HE}; e.g.][]{BastianHe,Chantereau:16}.  However, since the NeNa reaction is very efficient, a large fraction of material in the massive star core does have the correct Na pattern before an extreme He enhancement is produced early in the life of the star. Thus, it is possible to reproduce the observed $\Delta$ Y if some mechanism is able to increase the mass loss at critical rotation and halt self-pollution before large amounts of He is injected in 2P stars (i.e. if the core material can be accessed earlier than models predict).  This would, however, introduce a high degree of fine tuning.

\item  Discs where 2P stars are forming must detach at a certain stellar mass/age (which varies from star to star depending on its initial mass and metallicity) to avoid pollution by He-burning products, i.e. a strong increase of C and O not allowed by observations.

\item Massive stars ejecta are also Li free, so one must invoke some degree of dilution with unprocessed material to reproduce observations (see \S~\ref{SEC:LI}). 

\item Rotating massive stars would coexist with the supernovae from single stars as well as with other massive stars. Hence, it is not clear how their discs can survive in the crowded central GC regions \citep[e.g.][]{Renzini:15}.

\item  2P abundances would have necessarily continuous distribution. The photometric and spectroscopic discreteness observed in
some clusters cannot be readily reproduced by massive stars (Krause et al. 2016).
\end{itemize}

\subsubsection{Very Massive Stars (VMS)}

\citet{Denissenkov:14} envisioned a scenario where the most massive stars in the young cluster sink to the centre as a result of dynamical friction. Shortly after they reach the centre, the massive stars undergo multiple collisions with each other in a runaway process eventually forming a very massive star. VMSs with masses $\sim$ 10$^{4}$ $\msun$  are predicted to be fully convective with luminosities close to the Eddington limit, allowing for a significant mass loss. Below are some important constraints on VMS as the polluting stars.

\begin{itemize}

\item By the end of their MS lifetimes, VMSs are expected to reach very high He fractions, that would contradict the observed limits of  $\Delta$Y  in GCs today (\S~\ref{SEC:HE}).  Hence, in order to stop the overproduction of He, it has been suggested that VMSs fragment (soon after it formed), when only a small fraction of H was transformed into He. Thus, hot H-burning should occur only for a limited amount of time during the MS evolution on a VMS to reproduce the observed $\Delta$ Y distribution; e.g. until the Y has increased to Y$\sim$ 0.4.

\item While the observed anti-correlations and the Mg isotopic ratios --contrary to the case of AGBs and FRMSs -- are nicely reproduced, VMS nucleosynthesis cannot account for the observed Li (\S~\ref{SEC:LI}).  Therefore, dilution is also required in this model.

\item Only stars with masses in the mass range between 2 $\times$ 10$^{3}$ - 10$^{4}$ \msun~have central temperatures that provide the observed  GC light element anomalies up to Mg (e.g.; Prantzos, Iliadis \& Charbonnel 2017).

\item VMSs have not been observed and their existence is still highly speculative. Also, due to the relativistic conditions required to model them, which in general has not been included in stellar evolutionary codes, their evolutionary and nucleosynthetic yields are also highly uncertain.

\end{itemize} 

  \begin{marginnote}[]
\entry{SDU}{Second Dredge-up}
\entry{TDU}{Third Dredge-up}
\entry{HBB}{Hot Bottom Burning}

\end{marginnote}

\subsubsection{AGB Stars}

Processed material with some of the observed 2P chemical composition can be provided by intermediate-mass ($\sim$5 - 6.5 $\msun$) AGB stars through a complicated interplay of nucleosynthesis and mixing episodes, namely the SDU, the TDU, and HBB \citep[e.g.][]{Karakas:14}.  Contribution by lower mass AGBs should be avoided because AGBs less massive than $\sim$ 3.5 $\msun$ would release into cluster ejecta with enhanced C+N+O content\footnote{Surface C+N+O enhancements is also predict for rotating AGB stars more massive than $\geq$ 4 \msun \citep{Decressin:09}, contrary to observations of the majority of GCs.}.

During the SDU the convective envelope extends into the H-exhausted region and mixes to the surface mostly He and N from the CNO cycling. Ashes from He burning nucleosynthesis (mostly C and O, as well as Na and Mg) are eventually transported from the interior to the surface by the TDU, leading to an increase of the total C+N+O in the ejecta.  Following each TDU episode, the H-burning shell is re-ignited until the next instability of the He-burning shell develops. This exchange of power between H- and He-burning shells along with the associated TDU episodes occurs many times during the AGB phase and the overall changes in the surface abundances of AGBs stars caused by TDU episodes strongly depends on mass, metallicity, mass-loss, etc.

Intermediate-mass stars also  have envelopes that can reach very high temperatures (up to $\sim$ 100 MK, with the maximum temperature reached is a function of the AGB mass) to activate hot H-burning.  This process is known as HBB.  As a result, the envelope is exposed to regions where hot H-burning takes place, until the temperature at the base of the convective envelope drops below $\sim$20 MK (because of the mass loss which removes the envelope) at which point HBB is not longer supported.

 \begin{textbox}[!hb]
 \section{Comparison of AGB Model Yields}\label{BOX:AGB}
 The chemical evolution of AGB star models greatly depends on the adopted input physics. 
Different  treatments for convection and mass loss recipes lead to variations of the HBB or TDU efficiency (among others) in the AGB models, indirectly changing the chemical yields. As a result,  {\em "the predictive power of AGB models is still undermined by many uncertainties''} \citep[][]{Ventura:05}.

Models based on the mixing-length theory (MLT) of low convective efficiency fail to reproduce most of the observed chemical anomalies \citep[e.g. ][]{Fenner:04,Doherty:14}. In particular, they predict HBB temperatures that are too low to allow for efficient ON processing, i.e. AGBs produce too much Na and they do not provide large O depletion. Also, Mg and Al are positively correlated in the yields. 2P stars would also show an increase in the total CNO, which contradicts observations \citep[][]{Ivans:99}. 

Full spectrum of turbulence (FST) models are, compared to MLT case, more consistent with observations on MPs. FST model for turbulent convection results in a large convection efficiency, which translates in a very strong HBB \citep[e.g.;][]{Ventura:05,Ventura:05a}. Higher temperatures are reached at the base of the convective envelope and stars evolve to higher luminosities with respect to the MLT case. As a consequence of the high luminosity and larger mass-loss, they undergo a limited number of thermal pulses, so that the impact of TDU in changing the surface composition is limited. However, the lack of TDU in the FST models also limits the amount of Na that can be produced in AGB stars with M$\geq$ 5 $\msun$, which reach temperatures so high that sodium is destroyed, providing a negative sodium yield. The theoretical yields may be reconciled with the observations only if we assume that the (uncertain) cross section of the main channel of sodium destruction is a factor of $\sim$2-5 lower than the recommended values \citep[][]{Ventura:06,Dantona:16}. Finally, in the FST  case, the magnesium isotopic ratios are expected to exceed (by far) unity in the more massive stellar models (M $\geq$ 4 \msun), in contrast to what is observed \citep[][]{Yong:03}.
 This problem is shared by the MLT model.

\end{textbox}

A summary of the ability of AGB stars to match observed MP abundances is as follows:

\begin{itemize}

\item Pollution from AGBs qualitatively reproduces some of the light element variations observed in 2P stars. However, it is not possible -- without some modifications to the main physical inputs and relevant cross sections -- to obtain 
simultaneous O depletion and Na enrichment and keeping the C+N+O sum constant in AGB yields,
as required by observations \citep[e.g.][]{Sneden:00coll,Charbonnel:16,Marigo:17}.
Indeed, the composition of the material ejected by AGBs through winds critically depends on what mechanism (either TDU or HBB) dominates. The net effect of TDU is the mixing of He-burning products to the surface, in particular, C, Ne and O. The HBB destroys O and produces Na by p-captures on the dredged-up Ne (note that the surface Na abundance first increases during the SDU). At the temperatures required to destroy Mg ($\sim$ 100 MK), Na is destroyed again \citep[e.g.][]{Denissenkov:03}. Thus, without the Ne dredged-up by TDU and converted into Na by HBB, low values of O in the ejecta would lead necessarily to low Na for very high temperatures \citep[e.g.][]{Denissenkov:03}.  Na production can be increased by invoking an efficient TDU to effectively replenish Na by dredged-up Ne. However, this would lead to an increase of the overall CNO sum that is not allowed by observations. Alternatively, Na destruction can be lowered by tweaking reaction rates \citep[][]{Renzini:15,Dantona:16}. 

\item The observed Li distribution is not reproduced and dilution with a large amount material characterised by the same GC pristine composition (same initial abundances and Fe) is needed \citep[e.g.][see \S~\ref{sec:dilution}]{Dercole:16}.  Dilution is also required in order to obtain the observed anti-correlations (e.g., Na-O).  As the cluster is $>30$~Myr before AGB stars evolve, it is not clear where this material would come from (see \S~\ref{sec:origin_of_dilution}).  The need to dilute AGB ejecta with unprocessed material also requires that material from the first massive stars exploding as SNe should be removed from the cluster, e.g. in order to avoid variable pollution from Fe-rich material resulting in [Fe/H] spreads.

\item  He-rich material is mixed into the surface via the SDU, whereas the TDU and HBB are responsible for changes in light elements.
Thus, He, Na and Al should not be strictly correlated in AGB yields \citep[e.g.][]{Charbonnel:16}\footnote{Even if some initial Na enrichment during the SDU is expected, Na production due to the burning of dredged-up Ne also contributes to the resulting Na abundance. Thus, an obvious correlation between Na  and He is not expected {\em a priori}.}. The He content of the ejecta is predicted to increase with stellar mass, and can reach He values  up to Y$\sim$ 0.38  in super-AGB stars \citep[e.g.,][]{Ventura:13}, less than that observed in some GCs.

\item Since the temperatures reached during the HBB are related to the envelope opacity and thus to the overall metallicity of clusters, the AGB model would naturally explain why the products of extreme nucleosynthesis (Mg depletion and Si and K production) are observed only in metal-poor clusters. However, it is not clear why many metal poor clusters do not show these trends.  The HBB temperature may be high enough to alter Si and K abundances in the most massive AGB models \citep[e.g.; ][]{Ventura:12}, however at such temperatures Na would be destroyed, i.e. 2P stars would have low Na abundance \citep[][]{Charbonnel:16}.

\item Low-mass AGBs could potentially be responsible for the star-to-star variations in C+N+O and s-processes observed in a handful clusters (\S~\ref{SEC:PECULIAR_GCS}). However, they cannot produce light-element variations themselves
(because of the competition between TDU and HBB).

\end{itemize}


\section{Theories for the Origin of Multiple Populations}
\label{sec:models}

\subsection{AGB Scenario}
\label{sec:models_agb}
AGB stars have been suggested to be the source of the polluted material, early on in the development of this field \citep[e.g.,][]{Cottrell:81}.  The ``AGB Scenario" is arguably the model that has gotten the most attention in the literature, and many aspects of the model have been included in other scenarios, even those that use different enrichment sources.  Hence, we begin by discussing this model.

\subsubsection{Basic Scenario}

The model envisions the formation of a massive cluster with a single age and abundance pattern (i.e, an SSP), representing a first generation (FG) of stars.  The feedback from high mass stars and the associated SNe clear any remaining gas from within the cluster, hence all enriched material from the high mass stars and SNe are lost from the cluster (this is required to avoid Fe-spreads).  After $\sim30$~Myr, stars from the FG begin to evolve through the AGB phase of stellar evolution, and the winds of these stars, due to their low velocity ($\sim10-30$~km/s - \citealt{Loup:93}), are not able to escape the cluster, so a reservoir of polluted gas begins to form in the cluster.  This material cools and sinks towards the cluster centre, and once a critical density is reached a second generation (SG) of stars begins to form out of this material \citep[e.g.,][]{Dercole:08,Bekki:17}.   Early versions of the model had the second generation forming more or less continuously until star-formation was truncated due to the onset of 'rapid' Type-Ia SNe which would clear the cluster of any remaining gas, at an assumed age of $\sim100$~Myr.  After the sub-populations within GCs were found to be largely discrete (e.g., M4 - \citealt{Marino:08M4}), the model was refined by invoking multiple discrete bursts between the onset of AGB stars ($\sim30$~Myr) after the formation of the FG and when Type-Ia SNe began \citep[e.g.,][]{Dercole:16}.  

\begin{marginnote}[]
\entry{First generation stars (1G)}{In models of MP formation, stars of the first generation}
\entry{Second generation stars (2G)}{In models of MP formation, stars of the second generation that show the anomalous chemistry.}
\end{marginnote}

\begin{figure}[!tb]
\includegraphics[width=0.75\columnwidth]{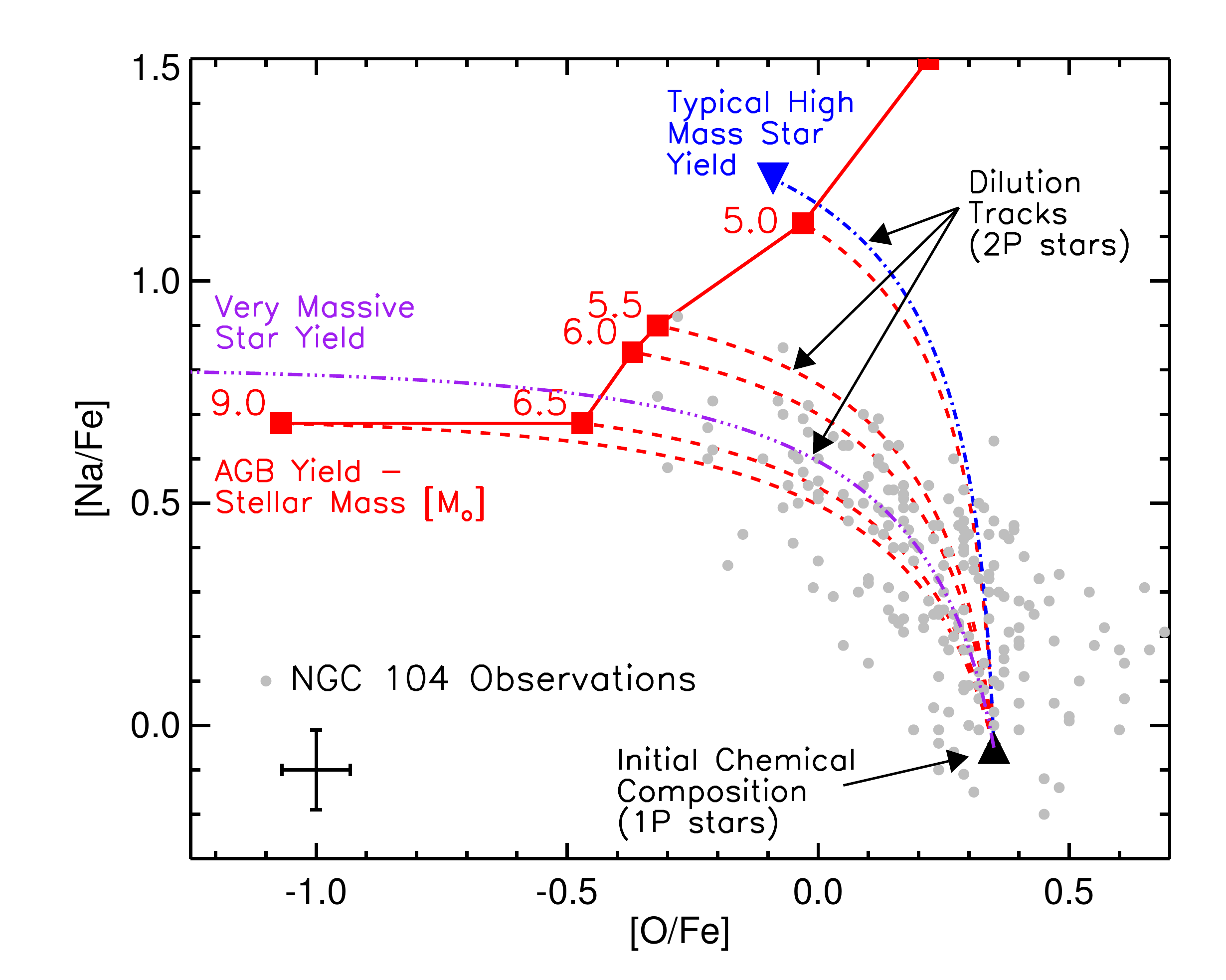}
\caption{An illustration of a dilution model.  The yields of suggested polluter stars are shown: AGB yields (from \citealt[][]{Dercole:10}) are shown with red squares for different masses (although note that other AGB yields do not show significant Na enhancement - \citealt[][]{Doherty:14}); typical high mass star ($\sim20$~\msun) yields are given with a blue upside-down triangle (from \citealt[][]{deMink:09}) and very massive star ($\sim5\times10^4$~\msun) yields are shown (off to the left of the panel - from \citealt{Denissenkov:14}). Dilution models use these yields and then dilute them with gas that has the initial chemical composition (i.e. that of the 1P stars). This leads to dilution tracks where the 2P stars are located.  All suggested pollution mechanisms require dilution (to various degrees) to explain the observed chemical abundances (i.e. He and Li).  Also shown are data from NGC~104 from the compilation of \citet[][]{Roediger:14}.}
\label{fig:dilution}
\end{figure}

It is worth noting that all AGB models to date do not produce a Na-O anti-correlation, but rather a correlation.  In order to reproduce the observed anti-correlations, this scenario requires the (re)accretion of large amounts of pristine material (i.e. material that shares the same abundances as the FG stars) from the surroundings, i.e., dilution of the AGB yields with material that matches the initial chemical composition of 1P stars is required.  In Fig.~\ref{fig:dilution} we show the basic idea of a dilution model.  Combining the yields from the polluting stars (e.g., AGB stars) with material that matches the 1P stars, dilution tracks can be created to explain the run of chemical abundances observed within clusters, where a 2P star's position is governed by the relative amount of processed material (i.e. AGB yields) and diluting material (1P chemistry) used to form the star. 

This accreted material material is then mixed with the AGB ejecta and forms SG stars, known as dilution, hence the SG of stars would have different Na-O abundances ranging from the pure yields of AGB stars to those of the FG. An additional problem for yields of AGB stars is that in the mass range of $\sim4-9$~\msun\footnote{AGB stars of lower masses are generally disregarded as contributing to the formation of the SG as they do not conserve the C+N+O sum, which contradicts 
observations \citep[e.g.,][]{Ivans:99}} some models provide the Na-enrichment and O-depletion required to match observations  \citep[e.g.,][]{Ventura:09}, whereas other calculations have found that AGB stars are not able to produce the Na-enrichment required \citep[][see \S~\ref{BOX:AGB}]{Doherty:14}.  Additionally, the latter models find that the C+N+O sum is not kept constant at any mass for AGB stars, in conflict with the observed properties of MPs in most clusters.

\begin{marginnote}[]
\entry{IMF}{Initial Mass Function of Stars}
\end{marginnote}

An important aspect of this - and most other - models, is that they can only produce a small fraction of the total cluster mass in 2G stars.  This is due to the stellar IMF of the 1G of stars, which only has a small fraction of its total mass in stars in a specific mass range that can produce material to pollute/enrich the 2G of stars (i.e. $f_{\rm enriched,initial} \sim0.02-0.1$).  In order to obtain the observed fractions of primordial and enriched stars (\fenrich$=0.4-0.8$) the model needs to assume that GCs lose substantial fractions of their initial population of stars (1G stars), often up to 95\% of their initial masses while retaining all/most of the 2G stars\footnote{Heavy mass loss is also required by the FRMS scenario \citep[e.g.;][]{Schaerer:11}.}. This will be further discussed in \S~\ref{sec:internal_mass_budget}.

In the model envisioned by \citet{Dercole:08} the gas coming off of AGB stars is able to rapidly cool, mix with material (possibly accreted) with the same chemical abundance pattern as the 1G stars, fall to the centre of the cluster, and subsequently form a 2G of stars.  However, it is not clear whether such material would be able to cool and remain in the cluster.  For example, if the heating of a population of x-ray binaries is included in the simulation, the gas is unable to cool, and instead flows out of the cluster.  \citet{Conroy:11} have shown that  the Lyman-Werner photon flux of stars of the 1G is high enough to not allow the gas to cool and sink to the cluster centre, until an age of $200-300$~Myr, delaying the formation of a 2G of stars for a much longer period of time.  Such a time delay would be a severe problem for the AGB scenario, as even under optimistic model yields, the C+N+O sum would not be conserved for AGB stars at this mass. \citet{Conroy:11} have also shown that due to the cluster's motion within the galaxy, Bondi-Hoyle accretion onto the cluster is expected to be very inefficient, and the authors suggest that clusters can retain a relatively large fraction of their initial gas mass ($\sim10$\%) in order to sweep up the interstellar medium (ISM) in order for the cluster to have the necessary primordial gas for dilution.  This again, ignores the role of heating from x-ray binaries and other mechanisms not included in standard simple stellar population models, whereas if such sources are included clusters would be expected to be gas free, which is a substantial problem for this model (see \S~\ref{sec:ymcs}).  It is also not clear that the material accreted from the surrounding galaxy would match the abundances of the 1P stars to the necessary precision imposed by the lack of Fe spreads in most clusters. 

One of the features of AGB stars that make them promising candidates to supply the enriched material is the fact that they can burn H at higher temperatures than main sequence massive stars, the exact ranges depends on metallicity and mass of the AGB star (see e.g., Fig.~8 in \citealt{Prantzos:07}).  This allows them to activate the Al-Mg burning chain, hence to deplete Mg and increase Al.  As discussed in \S~\ref{SEC:CN}, a minority of clusters show significant Mg spreads and most other potential polluting stars have difficulty producing the spreads without adjusting the nuclear cross-sections in an ad-hoc manner.  By including dilution, the basic AGB model (for some model yields) is able to quantitatively match the observed Na-O anti-correlation with GCs and qualitatively the increase in He.  On the other hand, it does not predict the correct abundance pattern of Li (as material processed through AGB stars should, to first order, be Li free) without invoking and fine tuning a specific mechanism to produce Li (see \S~\ref{SEC:LI}).

In summary, the basic AGB model, while conceptually simple, has a number of shortcomings that subsequent works have attempted to address.  This will be explicitly addressed in \S~\ref{sec:comparisons}.

\subsubsection{Alternative Versions}

In order to avoid the problems associated with dilution (i.e. accreting the material and the associated timing constraints), \citet[][see also ~\citealt{Renzini:15}]{Renzini:13} suggest that the yields of AGB stars may be very different from that predicted by current theoretical yields.  Due to the many parameters involved in estimating the yield of AGB stars (see \S~\ref{BOX:AGB}) there is significant freedom when adopting AGB yields.  The authors speculate that  perhaps the true yields of AGB stars result in a Na-O anti-correlation so that no dilution would be necessary, although without dilution it would be very difficult to match the Li abundance patterns.  Further work is needed to search the full range of potential parameter space of AGB model yields, but work so far suggests that AGB stars are not able to produce an anti-correlation of Na-O \citep[e.g.,][]{Marigo:17}.  However, if this was true, it would add an additional factor of $\sim2$ to the already strict mass budget problem (which is discussed in detail in  \S~\ref{sec:mass_budget}).  

It has also been suggested that ancient GCs may have formed embedded in larger dark matter halos, allowing them to hold onto a larger fraction of the material ejected from evolving stars \citep[e.g.,][]{Bekki:07, Trenti:15}.  If large/extended dark matter haloes were necessary to form MPs, then we would expect that only the oldest ($z_{\rm form} > 6$) GCs would be able to host MPs, as at lower redshift it would be increasingly unlikely to find a gas-rich dark matter halo that has not undergone significant star-formation (where Fe spreads would be expected).  The discovery of MPs in clusters younger than $8$~Gyr ($z_{\rm form} < 1$) argues against this type of scenario \citep{Hollyhead:17, FlorianSMC}.

\subsection{Fast Rotating Massive Stars (FRMS) and Interacting Binaries (IBs)}

Massive stars also undergo hot hydrogen burning in their cores, during the MS, and as such are also potential candidates to provide the enriched material needed to form MPs.  However, as this happens deep within the stars it is difficult  to bring up the enriched material to the stellar surface where it can be released into the GC intra-cluster medium.  Massive stars that are rapidly rotating can overcome this problem, due to rotationally induced mixing which can cause, in extreme cases, the stars to be (nearly) fully mixed.

\citet{Decressin:07a} and \citet{Decressin:07b} developed a scenario using FRMS as the enrichment source.  This scenario is similar to that of the AGB scenario, using the enriched material from a FG of stars to form a SG, but happens when the cluster is much younger ($<10-20$~Myr).  As in the AGB scenario, the ejecta of FRMS must also be diluted to match the observed abundance patterns (typical yields and dilution are shown in Fig.~\ref{fig:dilution}).  However, since the cluster is still young there is no need to bring the material from outside the cluster, as it is assumed that the cluster has retained a relatively large fraction of gas/dust left over from the formation of the FG.  The winds of the FRMSs then mixes with the left over gas and forms a SG of stars.  The FRMS scenario suffers from the same mass budget problem discussed for the AGB scenario \citep[e.g.,][]{Schaerer:11}.

FRMS naturally produce a Na-O anti-correlation and the enriched material can also be strongly enhanced in He, which helps explain clusters with large He spreads like NGC~2808. However, the high He yields may be a problem for more typical clusters with small He spreads \citep[e.g.,][]{Chantereau:16}.  FRMS are not able to activate the Al-Mg chain before the end of the MS, so are not able to explain the observed Mg spreads in some clusters without ad-hoc changes to the nuclear cross sections. 

\citet{Krause:13} further developed the FRMS scenario by exploring cases where a young GC may not be able to expel the left-over gas from the formation of a 1G of stars, even with SNe, allowing the cluster to remain embedded in it's natal GMC for $\sim20$~Myr.  The authors suggest that the decretion discs (i.e. equatorial discs forming from material that is thrown off the critically rotating star) might also accrete material from the gas rich intracluster medium, which would solve the dilution requirements.  

\begin{marginnote}[]
\entry{GMC}{Giant Molecular Cloud}
\end{marginnote}

\citet{Charbonnel:14} presented a variant on the FRMS scenario in order to solve the mass budget problem (see \S~\ref{sec:mass_budget}).  Here, the first generation of stars forms with a top heavy stellar IMF (i.e., only stars that would not be alive today) and the second generation would consist mainly of low mass stars.  In this model, stars with ``primordial composition" (i.e. 1P stars) would be actually second generation stars that formed primarily from material left over from the first generation.  Such a model can be tested through carbon isotopic ratios of MS stars.

Another way to release enriched material from the cores of massive stars into the intracluster medium is through binary interactions.  \citet{deMink:09} modelled a binary interaction between a 20 and 15~\msun\ star and investigated the yields of the expelled material.  They found that the 20~\msun\ star shed about 10~\msun\ worth of material due to the interaction, and that the yields matched the observed trends in GCs (i.e. Na-enriched, O-depleted, etc).  While the overall trends and correlations of the yields should apply to most massive stars, the exact yields depend on a number of parameters, e.g., the time of interaction (i.e. stellar evolutionary state), total mass of the stars and the mass ratio of the stars.  Hence, interacting binaries have the benefit of potentially explaining the observed variations from cluster to cluster, but have difficulty matching the discreteness of abundance ratios found in many sub-populations. 

A potential problem of scenarios that operate in the first few Myr of a cluster's life, is that after 3-8~Myr (depending on the cutoff mass for SNe), core collapse SNe begin to explode.  The retention of just a small amount of this material will result in Fe spreads that are in conflict with observations \citep{Renzini:08}.  Hence, processes that are limited in time to the epoch before the first core collapse SNe may need to be required in such models.

\citet{Szecsi:18} proposed a variant on this scenario, where a 2G of stars form in shells around high mass ($150-600$~\msun) red supergiant stars.  This scenario suffers from similar problems as the FRMS scenario (in terms of abundances, discreteness, and mass budget), but also is only expected to operate at low metallicity.  Since MPs are found in GCs of all metallicities ($-0.3 > [Fe/H] > -2.5$) this scenario could only apply to a subset of the known GCs.

\subsection{Early Disc Accretion Scenario}
\label{sec:EDAS}

\citet{Bastian:13} suggested an alternative model for MPs that did not invoke multiple epochs of star-formation.   Instead, it was driven largely by the constraints posed by YMCs (see \S~\ref{sec:ymcs}).  The model used the enriched material ejecta from high-mass interacting binary stars \citep{deMink:09} as well as the FRMS within the cluster to pollute low mass stars that formed at the same time as the high mass stars.  The authors suggested that low-mass ($<2$~\msun) stars may retain the protoplanetary discs around them for $\sim10$~Myr which would sweep up the enriched material as they passed through the cluster core (the authors also assumed that the cluster is mass-segregated from a very early age, so that the high mass stars are concentrated in the cluster centre).  The enriched material that was swept up by the discs would then eventually be accreted onto the host star.

While this scenario matches most observations of YMCs, it has a number of shortcomings as well (see \S~\ref{sec:comparisons}).  In particular, it requires that the accreting stars are fully convective (in order to mix the accreted material throughout the star) which in turn means that the accretion timescales are extremely short (1-3~Myrs - \citealt{Salaris:14,Dantona15:pms}).  This minimises the time that the mechanism could potentially work which effectively limits the amount of processed material that can be supplied and accreted.

\citet{Wijnen:16} ran hydrodynamical simulations to test this scenario, placing a realistic protoplanetary disc in a ``wind" of material (i.e. the ejecta of interacting binary stars, where the ``wind" refers to the disc moving through the intracluster ISM).  They found that while the disc did indeed accrete material from the ISM, the accreted material had little or no angular momentum which caused the disc to rapidly accrete onto the star and disappear.  Without the disc no further accretion would be possible.  The authors found that this happened on a rapid timescale, $\sim10^4$~years, much shorter than the required $10^7$~years for the scenario to work.

\subsection{Turbulent Separation of Elements During GC Formation}

\citet{Hopkins:14} also put forward a potential origin of MPs that did not invoke multiple generations of star-formation within GCs.  In his scenario, MPs would be the result of cloud physics during the earliest phases of GC formation.  In extremely turbulent environments, like those in progenitor clouds of GCs, large dust grains can become aerodynamic and begin to move separately from the gas and small dust grains.  Large resonant fluctuations in the dust can then develop.  Within these overdense regions, dust will be over-represented, so any stars that form within such regions will be enhanced in the elements associated with large dust grains.  On the other hand, the gas and small dust grains (like Fe grains) will be more uniformly distributed.  In principle, this mechanism provides a natural and powerful way to separate elements in the early phases of GC formation. Since this mechanism depends on the level of turbulence, it would predict larger abundance spreads in more mass proto-GC clouds, consistent with observations.

However, as noted by the author, Na and O normally occur on the same dust grains, so such fluctuations would predict a Na-O spread but as a correlation instead of the anti-correlation seen in GCs.  Also, He is not affected by dust, so an additional mechanism would need to be invoked to explain the inferred He spreads in GCs.  Finally, any enhancement in an element in some stars would necessarily lead to a depletion of that element in other stars.   We would then expect that, starting from field star abundance composition, we would see more or less symmetrical spreads around the field star abundance.  Observations, however, show the scatter in a single direction from the position of where halo field stars lie (at a given metallicity).

\subsection{Reverse Population Order For GC Formation Scenarios}

In order to alleviate the mass-budget problem (which will be discussed in \S~\ref{sec:comparisons}), some authors tentatively investigated formation models where the abundances of forming stars move from 2P to 1P, as star formation within the cluster proceeds (e.g.; \citealp{Marcolini:09}, Pancino et al., in preparation).

The scenario outlined  \citet{Marcolini:09} envisions GC formation from gas enriched locally by a single Type Ia SN and AGB yields superimposed on an ambient medium pre-enriched by low-metallicity Type II SNe. The star formation of the proto-GC only takes place inside this region and stars born within the inner volume will be depleted in O and Mg (because of the single SN Ia) and enhanced in N, Na and Al abundances (due to AGB pollution). External to this volume can be found a region with the same composition as the proto-halo gas at the epoch of GC formation.  After a new generation of stars is born, associated SNe II begin to pollute and expand the inner volume, while mixing with the lower metallicity material from the external shell, i.e.  gas with pristine composition.
Hence, the [Fe/H] and the CNO sum remain constant during cluster evolution and the N-C and Na-O anti-correlations can be reproduced. The Al-Mg anti-correlation can only be reproduced assuming that AGBs produce more Al than predicted by models \citep[by a factor of $\sim$10-50; e.g.][]{Karakas:07}.

 In a following paper, the authors focus on other elements and achieve some success in reproducing the observed trends \citep{Sanchez:12}. Nonetheless, the dynamical feasibility of the scenario has not been probed with hydrodynamical simulations and severe assumptions need to be made on the Fe content of the ISM at the epoch of formation, as well as the 
 the size of the inner region where the inhomogeneous pollution by the SN Ia and AGBs is confined \citep[e.g.][]{Sanchez:12}. More importantly, this class of models require very peculiar stellar configurations that are not  expected at the present epoch \citep[e.g.][]{Conroy:12}. 

\subsection{Extended cluster formation event}

\citet{Elmegreen:17} have further explored a model put forward by \citet{Prantzos:06} that invokes the special conditions of galaxies or GMCs at high redshift (namely high density, turbulence and pressure environments) to foster the formation of MPs before the first SNe occurs ($<3$~Myr).  Here, a first generation SSP is born in the core of a massive, dense and turbulent GMC.  Due to the high stellar densities, high mass stars have their envelopes stripped (and rotating massive stars lose large parts of their envelopes through decretion discs) very rapidly, which (as discussed above) are expected to show many of the observed abundance anomalies.  This material mixes with that left over from the formation of the FG and forms subsequent generations.  Low mass FG stars are assumed to be ejected due to two mechanisms, the first is binary dynamics and the second is that the gravitational potential of the cloud core/cluster is rapidly varying as the gas within it (which dominates the potential) is moved due to stellar feedback.  It remains to be seen if the high FG mass loss rates (and required low SG mass loss rates) required are feasible.

\citet{Wunsch:17}, following on \cite{Tenorio:05}, have suggested that the winds released from massive stars can become so dense in a massive and dense young cluster, that they enter a catastrophic cooling regime and can collapse into the cluster centre.  Here, the material may mix with left over primordial material (i.e. dilute) and form a second generation of stars.  Hence, this is another mechanism (rather than stellar interactions) that can potentially make enriched material from massive stars available for further epochs of star-formation within a cluster.  This also suffers from the mass budget problem and would require large fractions of 1G stars to be lost.  \citet{Lochhaas:17} develop this model further in terms of chemistry and show that the model is not able to simultaneously account for the increasing enriched fraction and increasing chemical spread with increasing cluster mass (see \S~\ref{sec:trends_cluster_properties}).

As these scenarios invokes massive stars, we will include it in our comparisons with observations, in particular the abundance trends, with other scenarios that invoke massive stars (\S~\ref{sec:comparisons}).

A key aspect of this scenario is that it happens (and terminates) before the first SNe occurs within the proto-GC in order to avoid Fe spreads (similar to the FRMS scenario).  One potential problem with the scenario is that it takes high-mass stars some time to increase their He mass through nuclear burning, whereas this model starts using stripped material from the massive stars at $t=0$.  This may be ok for standard clusters with small He spreads (e.g., NGC~104) but it may be difficult to reproduce clusters like NGC~2808, which hosts a large He spread.

Finally, for the limited models available of interacting binaries and fast rotating mass stars, it is not clear that they will be able to provide the stochasticity (i.e., the specific abundance pattern - extrema, discrete sub-populations - for each GC) required to match the observations.  \citet{Elmegreen:17} suggest that sub-clumps may form within the proto-GC, and each sub-clump would have its own chemistry due to the exact chain of stellar interactions.  However, these sub-clumps would each be expected to be $>10^4$~\msun, where the stellar IMF is fully sampled, hence stochastic effects would be expected to be minimised.   The \citet{Wunsch:17} scenario suffers from the same problem.

\subsection{Very Massive Stars Due to Runaway Collisions}
\label{sec:vms}

\citet{Gieles:18} have developed a model for MPs that adopts VMSs ($>10^3$~\msun) as the origin of the processed material.  In this model, the proto-cluster undergoes adiabatic contraction due to gas accretion, increasing the stellar density and subsequently the stellar collision rate.  A runaway collision process can form a VMS, which releases hot-hydrogen burning processes through its stellar wind into the intra-cluster environment \citep{Denissenkov:14}.  This processed material mixes with pristine gas (i.e. gas with the same abundance pattern as the initial proto-cluster) and forms further generations of stars until the very massive stars burns out, or potentially explodes due to instabilities within the star.  Because the VMS can be continuously rejuvenated through stellar collisions, the amount of processed material ejected by the star can be several times the maximum mass of the star.  While this process leads to multiple generations of stars within the cluster, the expected age spread would be less than $\sim3$~Myr.

One major advantage of this model is that it predicts a super-linear scaling between the mass of the very massive star and the mass (or density) of the cluster.  This naturally produces the observed trend of increasing fractions of enriched stars (and potentially as well as the increasing spreads in N, Na, etc) as a function of GC mass.  This kind of model also does not violate the constraints from YMCs, and much of the expected abundance patterns also appear to match observations.

One of the major drawbacks of the model is that VMSs are still only theoretical, although the authors perform numerical simulations showing that under certain conditions (relevant for GC formation) runaway collisions are likely to take place, even when considering two-body relaxation and the strong stellar mass loss of the massive object due to its stellar wind.  This same process is expected to also be at work in clusters today, if they reach the required stellar densities.  Hence, it is not clear if the model can explain why NGC~1978 ($\sim2$~Gyr) hosts MPs while NGC~419 ($\sim1.5$~Gyr) does not, given their similar masses and radii.

\section{Comparing Predictions to Observations}
\label{sec:comparisons}

\subsection{Chemical Abundance Patterns}
\label{sec:chemical_abundance_patterns}

\subsubsection{The Need for Dilution}
\label{sec:dilution}

As discussed in \S~\ref{sec:models}, the suggested stellar sources for the origin of the polluted material have difficulties in reproducing some of the observed abundance trends.  For example, AGB model yields suggest that Na and O should be correlated, not anti-correlated.  Also, wherever nuclear processing of C, N, O, and Na takes place, the resulting material is expected to be Li free, whereas observations show that Li is constant or only slightly varying in GCs from star-to-star.  In order to address these problems, most models have adopted some form of dilution, i.e., that the enriched material produced by 1P stars is mixed with material that matches the chemistry of the 1P stars (referred to as ``primordial" material). 
Here we discuss the predictions of dilution in comparison with observations.  A basic illustration of a dilution model is given in Fig.~\ref{fig:dilution}.

\subsubsection{Li Variations}

Without dilution we would expect all 2G stars to be effectively Li free, as any material subjected to hot hydrogen burning will have its Li rapidly destroyed.  Observations show, however, that in some GCs Li is constant between 1P or 2P stars, or that it is depressed in 2P stars, i.e. anti-correlated with Na and correlated with O (see \S~\ref{SEC:LI}).  The amount of Li would then reflect the amount of diluted material included in the formation of 2P stars.  This assumes that Li is not produced by other processes.   In principle, AGB stars can produce some amount of Li through the Cameron-Fowler mechanism \citep{Cameron:71}, but this requires extreme fine tuning to match the observe Li variations/constancy (see \S~\ref{SEC:LI}).

\citet{Salaris:14} have pointed out difficulties in such dilution models.  Essentially, since the enriched material is expected to be Li free while only depleted in O, the spread in Li should always be larger than the spread in O.  However, for at least one cluster, NGC~6752, the spread in Li is smaller than the spread in O.  The Li spread (in relation to Na, O and other light elements) needs to be studied in other clusters, but if these results are confirmed this poses a major problem for all models that use high mass stars (i.e. $>15$~\msun) as well as AGBs.

There are tentative hints that the amount of Li variation is larger in higher mass clusters, similar to what is observed in Na, O, He and N. Indeed, in high-mass and metal-poor clusters, stars 
characterised by extreme composition (very high Na and Al enhancement) are also 
Li-poor \citep[e.g., NGC~1904, NGC~2808, NGC~6752, M~5, NGC~6397; see][and references therein]{Dorazi:15Li}.  The presence of a fraction of 2P stars with depleted Li abundance is surprsing, because 2P stars with an intermediate degree of chemical variations share the same Li abundance as 1P stars. If the light element anomalies are produced by nucleosynthesis in the interior of stars, this finding implies that some mechanism 
(either dilution or Li production by AGBs) should operate to restore the Li abundance of its initial value in 2P stars with intermediate composition without changing Li in extreme 2P stars. Again, such an interpretation requires extreme fine tuning.

\subsubsection{Quantitative Abundance Trends and the Need for Stochasticity}

While many studies have compared the observed abundance distributions of specific clusters to the yields of potential polluter stars, few have carried out a more general analysis including multiple elements and comparisons between clusters.  \citet{BastianHe} studied a sample of eight Galactic GCs that all had measurements of their Na-O anti-correlations as well as spreads in He based on HST imaging.  With the exception of NGC~2808, the authors conclude that the observed distributions (Na, O, He) were not in agreement with the predicted yields of AGBs, FRMS, interacting binaries, or very massive stars, even when dilution with primordial material was taken into account.  Specifically, based on the extent of the Na-O anti-correlations, large He spreads ($\Delta Y > 0.1$) would be expected in all  cases, whereas in most cases $\Delta Y_{\rm obs} < 0.05$. 

\citet{BastianHe} also considered ``empirical yields", i.e. adopting the observed Na-O anti-correlation and He spreads observed for a given cluster, and comparing that to the other GCs in the sample.  Surprisingly, even when using the ``empirical yields" a satisfactory fit for the other clusters could not be reached (even when controlling for metallicity).  The conclusion is that whatever the polluting source, it needs to produce a high degree of cluster-to-cluster variations in order to explain the observations.  Dilution of a fixed set of yields does not help in explaining the full set of observations. This argues against the  stellar sources normally considered (i.e., AGBs or massive stars) being the origin of the enriched material, as none can provide the necessary cluster-to-cluster variation.  However, the multi-modal abundance patterns within GCs suggest that for a given GC, the yield/dilution combination is quite uniform (i.e. taking on only a handful of values within the cluster).

It is beyond the scope of this review to quantitatively compare the yields of each proposed source with observed for each element, especially considering that most works to date have only focussed on one or two elements at a time (i.e. not testing whether the yields and required dilution that match a given element are able to match another).  We refer to the interested reader to e.g. \citet{Dantona:16,Prantzos:17}.

\subsubsection{The Origin of the Diluting Material}
\label{sec:origin_of_dilution}

For models that adopt massive stars as the origin of the enriched material ($>15$~\msun), it is assumed that a large reservoir of primordial material is left over within the cluster from the formation of the 1G of stars.

However, for models that invoke pollution from AGB stars, the origin of the diluted material is more difficult to explain.  Once core-collapse SNe from the massive stars in the 1G begin to explode, all material left over from the formation of the 1G is expected to be removed to large distances (i.e., unbound from the cluster).  Hence, the primordial material must then be (re)accreted from the surroundings.  This material must also avoid being contaminated with the material (e.g., Fe) from the SNe, or else Fe spreads would be expected in all clusters \citep[e.g.,][]{Renzini:15}.

\citet{Conroy:11} suggested that this material can be accreted from the host galaxy as the clusters orbit through the interstellar medium (ISM).  While accretion due to gravitational focussing is not efficient for the majority of cases, the authors found that if a reservoir of gas already exists within the cluster ($\sim10$\% of the stellar mass) it can sweep up material and the reservoir can grow.  However, \citet{Dercole:11} have shown that this near-constant accretion of new material, when coupled with the adopted AGB yields, will not reproduce the observed abundance distributions.  For the AGB model to work, the timing of the dilution needs to be very specific, with nothing being accreted (i.e. no diluting material present) when the most massive AGB stars are shedding their material, and an ever increasing amount of material being accreted after that, until the process is terminated, potentially by the onset of Type Ia SNe.

\citet{Dercole:16} have further developed the basic AGB scenario by placing the young GC inside a disc galaxy.   In the model, the SNe from the 1G of stars blows a hole in the surrounding ISM and eventually the expelled material is lost to the host galaxy.  They adopt the same basic scenario as \citet{Dercole:08} that the young cluster can retain the ejecta of AGBs and that this material can cool and form a 2G of stars within the cluster.  Eventually, the SNe blown bubble begins to close (as SNe become less frequent) and material from the galaxy fills the hole, some of which is then accreted back onto the cluster.  This scenario requires the surrounding material (out to 100s of pc) to be chemically identical to that of the FG stars within the cluster.  Additionally, this model does not take into account the motion of the cluster within the host galaxy, in particular the high velocity dispersion expected in young galaxies \citep[c.f.,][]{Kruijssen:15}, hence it is not clear that the gas would be accreted onto the cluster. Note that massive clusters ($>10^6$~\msun) in galaxy mergers today do not appear to be able to efficiently accrete material from their surroundings \citep[][]{Longmore:15,CabreraZiri:15alma}.

\subsubsection{Al-Mg Anti-correlation}

Interestingly, the presence of Al and Mg anti-correlated ranges among cluster stars is one of the strongest arguments against the FRMS scenario, as the temperature required to efficiently destroy $^{24}$Mg is reached in the core of massive stars only at the very end of their main sequence evolution \citep[][]{Decressin:07a}. As a consequence, a large increase (by a factor 1000) of the $^{24}$Mg(p,$\gamma$) reaction rate around 50 MK with respect to the nominal values is demanded to build the Al-Mg anti-correlation in the stellar core and even in that case Mg depletion would be associated with a strong He enrichment \citep[up to Y$\sim$ 0.8 after dilution with unprocessed material, see ][]{Chantereau:16}. Pollution from AGBs would in principle reproduce more naturally the observations, because both the depletion 
of Mg and the production of Al are sensitive to AGB metallicity, in the sense that more extended Al and Mg variations are expected at low metallicity, as observed \citep[][]{Ventura:16}.  
However, the resulting (anti-)correlations between Na, Mg, Al, and Si are greatly dependent on the mixing with He-burning material; e.g. because of the competition between TDU and HHB (see \S~\ref{SEC:NUCLEARREACTION}). 
Finally, the observed dependence of Mg depletion and Al production on metallicity  can be explained in the VMS scenario if the mass loss leads 
to the formation of smaller very massive stars at higher metallicities \citep[e.g.][]{Vink:11}.

\subsection{Discrete vs. Continuous Abundance Spreads}

In the majority of GCs, 1P and 2P stars are observed to be distributed continuously in the Na-O plane. However, a number of studies revealed that the O, Na, and Al abundances of different sub-populations are clustered around certain values \citep[e.g.][]{Marino:08M4, Lind:11,Carretta:142808,Carretta:15N2808}. Nonetheless, the evidence of multimodality from high resolution spectra is still sparse. On the contrary, C and N (and CN band strength) multimodality is almost universal among clusters with intermediate to high metallicity (e.g. \citealp{Norris:87})\footnote{The bands of bi-metallic molecules like CN are weak in metal-poor GCs, because their strength has a quadratic dependence on the metallicity.}.  HST photometry, in particular when including UV filters, also shows largely discrete RGBs and MSs in some cases (see \S~\ref{SEC:PHOTOMETRY} and Fig.~\ref{fig:uv_hst_legacy}).  These findings  indicate that the spectroscopic Na-O distributions may also be made up of discrete groups of stars, but that errors have blurred the distinction between the groups, causing the distribution to appear continuous \citep[e.g.][]{Carretta:13AL,Carretta:15N2808}.  

 The observed discreteness between two (or more) subpopulations would disfavour formation scenarios based on accretion onto pre-existing stars (e.g., EDA scenario) or 2P stars being born within the disk of fast rotating massive star (e.g., FRMS scenario). Such processes would result in a continuous range of abundance variations rather than the discrete distributions demanded by the observations.

\subsection{Radial Distributions, velocity dispersions and binarity}

The evidence of a more centrally concentrated 2P (see \S~\ref{sec:spatial_distributions}) is in qualitative agreement with most of the proposed scenarios. Also, the higher incidence of binaries with 1P composition \citep[][]{Dorazi:10,Lucatello:15} would again be consistent with a 2P preferentially found towards cluster inner regions.  For example, in the \citet{Dercole:08} scenario, where the AGB ejecta form a cooling flow and rapidly collect towards the cluster centre, forming a concentrated 2P.  The system starts with more concentrated 2P stars, as the cluster evolves, the 1P and 2P stars mix.  The long-term dynamical evolution of the different sub-populations with initial spatial segregation allows for efficient mixing in the innermost regions, where the local two-body relaxation time scale is shorter,  potentially erasing any initial differences between subpopulations on a relaxation timescale \citep[e.g.][]{Vesperini:13}. 

However, there are a handful of exceptions to this general behaviour, with the 2P stars {\em less} centrally concentrated than 1P stars \citep[][]{Larsen:15m15,Vanderbeke:15,Lim:16}.  While differences in mass between the 1P and 2P stars due to He variations offers a potential explanation, the required He spreads are much larger than can be accommodated by the observations \citep[][]{Larsen:15m15}. 

Different formation models may leave unique kinematics imprints  imprints that would allow to distinguish between various scenarios \citep[i.e.  different subpopulations showing different flattening; e.g.][]{Mastrobuono:13}. In this regard, the differential rotation of subpopulations provides precious insights, as such an observational property may survive the long term dynamical evolution of old GCs and would allow us to distinguish
different formation scenarios \citep[e.g.][]{Bellazzini:12,Vincent:15,Cordero:17}.

\subsection{Mass Budget Problem}
\label{sec:mass_budget}

A difficulty of all the proposed self-enrichment scenarios that was quickly realised was that since the enriched populations within GCs was equal to, or larger than, the primordial population (i.e., \fenrich$>0.5$), there simply would not be enough material processed through 1P stars to explain the number of 2P stars if standard stellar IMFs are adopted \citep[e.g.,][]{Prantzos:06}.  This is known as the ``mass budget problem".  For example, for a standard IMF, only $\sim7$\% of the stellar mass in a 1G is in stars with masses between $5-9$~\msun (i.e. stars that pass through the AGB phase often associated with the AGB scenario). Low mass ($<0.8$~\msun) stars, on the other hand, make up $\sim40$\% of the initial mass fraction.  For a typical GC, 2P stars represent $\sim67$\% of cluster stars, while 1P stars make up the remaining $\sim33$\%.  Assuming that 100\% of the mass of every AGB star gets used to make 2P stars (an extreme assumption) and that the 2P has a standard IMF, AGB stars can only account for $4-5$\% of the population of 2P stars.  If we assume that, on average, 50\% of the mass of each 2P star comes from diluting material, then AGB stars can account for $8-10$\% of the 2P stars.

The commonly invoked solutions to this problem have been 1) to apply an ad-hoc limit to the mass range of 2G stars to $<0.8$~\msun\, i.e. the mass range observed in GCs today (giving a factor of $\sim2.5$, i.e. accounting for $\sim20$\% of the needed mass),  and 2) to assume that the number 1G of stars was much larger when the cluster formed, and that $\sim90-95$\% of them have been lost during the evolution of the cluster.  These lost stars would then populate the field of the host galaxy.

For the AGB senario, \citet{Dercole:08} and \citet{Conroy:12} estimate that GCs must have been at least $10-20$ times more massive than observed today.  This is expected to be even $8-25$ times in the FRMS scenario.  \citet{CabreraZiri:15alma} discuss this problem in detail and conclude that under more realistic assumptions the problem may be factor of $2-3$ times worse than previously suggested (i.e., requiring clusters to have been $\sim30-60$ times more massive at birth than presently).  This is a basic prediction of these scenarios that can be directly tested observationally.

\subsubsection{Internal Mass Budget Problem}
\label{sec:internal_mass_budget}
We refer to the ``internal mass budget problem" to mean the relative numbers of enriched and primordial stars within GCs.  For a standard stellar IMF, only a small fraction ($2-8$\%) of the 1G mass is processed through a given stellar type (e.g., AGBs, FRMS, IBs) and released into the intracluster medium, even for optimistic yields.  However, the present day observed \fenrich~for clusters is $40-90$\% \citep[e.g.,][]{Milone:17}, and a significant amount of processed mass is needed for each of the enriched stars.  The standard solution to this problem is to assume that GCs were $10-100$ times more massive at birth than they are currently, and that, since the 2G stars are thought to be born more centrally concentrated, a large fraction of the 1G stars were lost during their evolution.  

\citet{Vesperini:10} have simulated the evolution of such a cluster in a Galactic-like potential and found that, in principle, with the right selection of parameters, such extreme mass loss can be reproduced with numerical models.  However, in order to obtain such extreme mass loss, the authors needed to assume that GCs began their lives tidally limited and mass segregated, so that they expand due to stellar mass loss and lose stars to the galaxy over their tidal boundaries. The clusters would then start their lives with effective radii of 10s to 100s of pc (depending on the strength of the tidal field at birth), although it has not been demonstrated that such clusters would resemble the observed Galactic GCs after $\sim10$~Gyr of evolution. Present day GCs and YMCs have much smaller effective radii, with means around $\sim3$~pc \citep{Harris:96,Larsen:04}. Additionally, it is not clear that such a mechanism would work in environments with weaker tidal fields (that display similar \fenrich\ as Galactic GCs) like that of GCs in the LMC/SMC or Fornax dwarf galaxy.

\citet{BastianLardo:15} and \citet{Milone:17} both looked at the \fenrich\ as a function of the Galactocentric distance.  If large fractions of 1P stars are lost due to the tidal field, even in the case of a tidally limited and mass segregated initial cluster conditions, there would be a strong expected relation between \fenrich\ and the Galactocentric radius \citep[see][]{BastianLardo:15}.  However, \fenrich\ was not found to depend on the Galactocentric distance (or orbit), in contradiction with predictions from scenarios that invoke heavy mass loss.  \citet{Milone:17} have found that \fenrich\ is a strong function of present day GC mass, with higher mass GCs having larger \fenrich\ (see Fig.~\ref{fig:correlations_with_mass}).  This trend is opposite to that which would be expected if GCs underwent large amounts of mass loss.  Higher mass clusters are expected to lose a lower fraction of their mass during their evolution, hence they should have enriched fractions closer to the primordial value.  

\citet{Kruijssen:15} estimated the mass lost from GCs forming and evolving in a cosmological context, and found that massive GCs (with initial masses $>5\times10^5$~\msun) are only expected to lose a relatively small fraction of their initial masses (i.e., potentially being a factor of $\sim2-4$ more massive than currently seen).  This is largely in agreement with non-MP driven estimates of mass loss from Galactic GCs \citep[e.g.,][]{Kruijssen09} and constraints from the form of the lower mass function in clusters (which is sensitive to mass loss - e.g., \citealt{Webb:15}.)

\subsubsection{External Mass Budget Problem}
\label{sec:external_mass_budget}

We refer to the ``external mass budget problem" to mean the number of primordial stars in GCs relative to that of the host galaxy.  This is linked to the internal mass budget if one adopts models where large fractions of 1P stars are lost to the field.  In principle, one should find an excess of 1P stars in the halo that came from GCs, at the position of the donor GCs in phase space (i.e. position, velocity and/or metallicity; e.g., \citealt{Schaerer:11}).

The number of GCs found (per unit galaxy mass or luminosity) is known to be high in some dwarf galaxies \citep[e.g.,][]{Larsen:12}.  It becomes even higher at low metallicity (e.g., [Fe/H] $< -$1 dex) when GCs and field stars of the same metallicity are compared \citep[e.g.,][]{Harris:02}.  \citet{Larsen:12} have exploited this observation to place some of the strictest constraints on the origin of MPs to date.  The authors counted the number of 1P stars in GCs in the Fornax Dwarf galaxy below [Fe/H]=--2 dex and compared that to the number of stars observed in the field in the same metallicity range.  They found that GC stars made up $\sim20-25$\% of the stars in this metallicity range.  Even if all stars in this metallicity range formed in clusters, this would mean that these GCs could have only been a factor of 4 or 5 more massive than they currently are, in contradiction with the requirements of models requiring large mass loss.  \citet{Larsen:14WLM} have extended this kind of study to the dwarf galaxies WLM and IKN and found similar results, showing that this is a common phenomenon and not linked to the specific evolutionary history of the dwarf galaxy host \footnote{In fact, the high specific frequencies observed in many dwarf galaxies \citep[e.g.,][]{Harris:13} argues against these heavy mass loss scenarios, assuming that GCs in dwarfs also host MPs (i.e. that MPs are ubiquitous).}.

\citet{Khalaj:16} have suggested that, in the context of the FRMS scenario, that the expulsion of gas (left over from the formation of the 1G and 2G stars) from the young GC could unbind large fractions of stars from the cluster at high velocity.  If the stars leave with a large enough velocity they could potentially leave the young galaxy all together.  Note that this solution would not be applicable to the AGB scenario, as the cluster would already be gas free when the AGB stars begin to evolve.  While possible, observations of YMCs today, do not support the idea that gas expulsion leads to large mass loss within clusters \citep[c.f.,][]{Longmore:14}.

\subsection{Trends with Cluster Properties}
\label{sec:trends_cluster_properties}

As discussed in \S~\ref{sec:age_mass} the present day mass of a GC is directly linked to 1) the fraction of enriched stars present and 2) the extent of the  abundance spreads in N, Na, O and He (see Fig.~\ref{fig:correlations_with_mass}), with higher mass clusters havng larger enriched fractions and larger spreads. Assuming that the yields of the polluting source (e.g., AGB, FRMS, VMS, etc) are not dependent on the GC properties, this is difficult to explain in the classic scenarios as stellar yields (for a fully sampled IMF) should provide a constant amount of enriched material per unit stellar mass.  

The increasing fraction of enriched stars at higher masses is contrary to the expectations of scenarios that invoke heavy mass loss to obtain large (present day) fractions of enriched stars (e.g, AGB or FRMS scenarios - see \S~\ref{sec:models} \& \S~\ref{sec:trends_cluster_properties}) as it would require higher mass clusters to lose larger fractions of their mass  (i.e. large numbers of 1P stars), opposite to that expected from basic dynamical considerations \citep[e.g.,][]{Kruijssen:15}.  Additionally, if GCs did lose large fractions of their initial masses, it would be extremely difficult to maintain these strong correlations with cluster mass (e.g., \citealt{Schiavon:13}).  It is also unexpected that higher mass clusters should show larger abundance spreads.  While in principle they may hold onto more of the processed material, they also should accrete/retain more primordial material (i.e., diluting material).  Additionally, models already assume that all of the processed material is used in the formation of 2P stars (i.e., 100\% star-formation efficiency of the processed material).

While it is difficult to reconcile models to these observations, it is worth noting that no model put forward to date is able to account for both the fraction of enriched stars as well as the extent of the variations as a function of cluster mass at the same time.  This is because, for the polluting sources suggested, the amount of enriched material produced is fixed per unit mass.  The model can use that enriched material to either create larger abundance spreads (i.e., putting more of it in 2P stars) are use it to create more 2P stars (i.e. increasing f$_{\rm enriched}$), not both.  The conclusion reached is that the enrichment mechanism must depend on the mass (or density) of the host cluster. 

\subsection{Constraints from Young Massive Clusters (YMCs)}
\label{sec:ymcs}

One of the major discoveries made by the HST was that stellar clusters with masses and densities rivalling (and in some cases, greatly exceeding) GCs are still forming in the local Universe \citep[c.f.,][]{PortegiesZwart:10}.  These YMCs are commonly referred to as proto-GCs as they have similar properties to that expected for the present day GCs when they were young \citep[e.g.,][]{Kruijssen:14}.  Due to their proximity and relative brightness, we can use YMCs to test the scenarios for the formation of GCs and the MPs within them.  The properties of YMCs themselves are discussed in \S~\ref{sec:ymcs_gcs}.

\begin{marginnote}[]
\entry{YMC}{Young Massive Cluster - a.k.a. young GC}
\end{marginnote}

While MPs have not been found to date within YMCs with ages of $<2$~Gyr \citep[e.g.,][]{CabreraZiri:16}, they can still provide useful constraints on the origin of MPs, as most theories put forward so far do not invoke any special physics present only in the early Universe.  Most theories simply invoke the gravitational potential of the young GC as being deep enough to hold onto expelled stellar ejecta.  Hence, even if YMCs are not the equivalent of proto-GCs, they can still be used to directly test predictions of the proposed scenarios.

\subsubsection{Constraints on Age Spreads within YMCs}
One of the key predictions of the AGB scenario is that clusters which are massive enough should be able to retain the ejecta of AGB stars and form subsequent stellar generations.  \citet{Larsen:11} studied the resolved CMDs of seven massive ($10^5 - 10^6$~\msun) young clusters in nearby galaxies, and while there were features in the CMDs that were not well described by a standard isochrone, age spreads (of the order of 10s of Myr) were also inconsistent with the observations. In one case, NGC~1313-379, an age spread could not be reliably ruled out.

Following on the work of \citet{Peacock:13}, \citet{Bastian:13ymcs} searched for evidence of ongoing star-formation within a sample of $\sim140$ YMCs with ages between $10$~Myr and $1$~Gyr and masses between $10^4 - 10^8$~\msun.  They searched for emission lines (i.e., H$\beta$ and O{\sc [iii]}) from the unresolved clusters and O-stars in the CMDs of the resolved clusters.  No clusters were found with evidence of ongoing star-formation.  \citet{CabreraZiri:14,CabreraZiri:16w3} took this analysis a step further by estimating the star-formation histories of two massive ($>10^7$~\msun) clusters in galactic merger remnants, NGC~34 (S1 - $\sim100$~Myr) and NGC~7252 (W3 - $\sim500$~Myr), using high S/N integrated optical spectra.  In both cases, the clusters were best fit by a SSP (i.e., no evidence of a secondary starburst was found).

At an age of $\sim2$~Gyr, NGC~1978 is the youngest cluster that shows evidence for MPs \citep[][]{Martocchia:18a}.  Due to its youth, it can be used to place tight constraints on age differences between the subpopulations.  \citet[][]{Martocchia:18b} were able to identify two populations on the SGB of the cluster with UV-photometry and then compared the positions of the stars in each population in an optical CMD.  In optical colours, the position of the stars along the SGB (essentially the vertical placement of the stars) is sensitive mainly to age (and not chemical anomalies).  The authors found an age difference of $1\pm20$~Myr between the populations, i.e. that they were coeval.

Taken together the constraints on age spreads in YMCs suggest that they are less than $10$ or $20$~Myr. This does not directly constrain the FRMS, VMS, or EDA scenarios, but does place severe restrictions on scenarios that adopt AGBs as the polluters, as the first AGB stars do not evolve until $30$~Myr after the 1G forms.

\subsubsection{Constraints on Gas and Dust Reservoirs within YMCs}
In order for a massive cluster to form a second generation of stars, it must be able to retain a significant amount of gas within it for an extended period.  
\citet{Longmore:15} used the predictions of the \citet{Dercole:08} model for multiple star forming events in the context of the AGB scenario, to show that the clusters should show extreme extinction in their inner regions, effectively being invisible in the inner $\sim3$~pc.  He notes that no such massive 'ring' clusters have been observed and that many massive ($>10^6$~\msun) clusters have been found with little or no extinction in the age range where the \citet{Dercole:08} models predicts that 2nd generation should be forming. \citet{CabreraZiri:15alma} used deep ALMA observations of three massive ($>10^6$~\msun) clusters in the Antennae merging galaxies with ages between 50 and 200~Myr to search for any gas within them.  Depending on the adopted conversion factor between the observed CO luminosity and total gas/dust mass, the authors could place upper limits of $<1-10$\% of the stellar mass is present in gas within the clusters.

Finally, \citet{Bastian:14frms} and \citet{Hollyhead:15} have studied a sample of young clusters ($<10-20$~Myr) with masses between $\sim10^4$ and $\sim10^7$~\msun\ to see how long clusters remain embedded in their natal gas cloud.  In contradiction to the predictions of the FRMS scenario of \citet{Krause:13}, who suggested that massive clusters should remain embedded for $\sim20$~Myr, observations showed that independent of mass (in the range studied) clusters were gas free within the first $2-4$~Myr of their lives, probably before the first SNe (for metallicities from $1/5^{\rm th}$ solar to solar). \cite{Whitmore:02} and \citet{Reines:08} studied the nearby starburst galaxies, NGC~4038/39 and NGC~4449, respectively, comparing  radio continuum measurements with optical HST colours/magnitudes.  Both works conclude that young massive clusters are largely gas free by an age of $7$~Myr, often considerably shorter.

We can conclude from these works that clusters are very efficient at removing (or consuming) any gas within them, from very young (few Myr) to very old ages ($>1$~Gyr).  This applies to very massive clusters, even if simple escape velocity arguments would suggest that they should be able to retain any gas within them.  For young clusters the Lyman-Werner flux within the cluster is expected to be very high \citep[e.g.,][]{Conroy:11} which will not allow the gas to cool sufficiently to collapse to the cluster centre, and the presence of x-ray binaries and other energetic sources (e.g., white dwarfs) and/or ongoing SNe appear to keep the cluster gas free throughout its lifetime.  Hence, models that invoke the potential well of clusters to hold onto enriched gas are not supported by observations.

\subsubsection{Globular Clusters in Formation}

A major advance in the field may come with the launch of the James Webb Space Telescope (JWST), as, in specific circumstances, it will allow us to peer into galaxies at the epoch of GC formation (i.e., $z>2$).  Some initial steps in this direction have already been taken by observing highly lensed galaxies at $z>3$ and their young GC populations.  \citet{Vanzella:17} studied a sample of compact GC-like objects in five highly lensed galaxies including rest-frame UV/optical photometry and spectroscopy from HST and VLT/MUSE.  Two of their objects, ID11 ($z=3.1169$) and GC1 ($z=6.145$) are particularly interesting due to their young ages ($<10$~Myr), small effective radii ($\lesssim50$~pc) and stellar masses ($1-20\times10^6$~\msun), which are expected for young GCs.  The estimated properties are also similar to those of YMCs forming in nearby galaxies, supporting the idea that YMCs are indeed the equivalent of young GCs.

It is difficult to draw conclusions from such a small sample, but future work on lensed samples (as well as JWST samples) offer a chance to study the populations statistics of young GCs.  If clusters are $10-30\times$ more massive when they form than they are currently, JWST would be expected to observe many clusters in excess of $0.5-1\times10^7$~\msun\ \citep[e.g.,][]{Renzini:17}.  Alternatively, in models for the evolution of GCs based on the observed properties of YMCs and the conditions expected to be experienced by the clusters throughout their lives (i.e., models not tuned to achieve severe mass-loss), only a handful of massive ($>0.5-1\times10^7$~\msun) would be expected in each host galaxy \citep[e.g.,][]{Kruijssen:15}.

\begin{summary}[Summary Points of the Comparison Between the Predictions and Observations]
\begin{enumerate}
\item The observed positive correlations between f$_{\rm encriched}$, extent of the abundance spreads, and cluster mass is directly at odds with scenarios that invoke large amounts of cluster mass loss in order to go from a cluster dominated by primordial stars to a cluster dominated by enriched stars.
\item This argues that the observed fractions are imprinted at birth, which essentially rules out all standard nucleosynthetic sources.
\item Quantitative comparison between the observed ranges of Na, O and He spreads with the predicted yields of suggested polluter stars shows that none (or any combination thereof) can match the observations.  While each cluster is unique in the details of its chemistry (requiring stochasticity in their formation) most clusters have He spreads that are much too small for the observed Na and O spreads.
\item Li is a problem for all scenarios, as it should be highly depleted in all material that is enriched in Na and He (and depleted in O), whereas observations do not show depletion to the predicted amounts (even including dilution).
\item YMCs, with properties similar to those expected for young GCs, are still forming today.  Studies have not found evidence for multiple star-forming epochs within the clusters, nor large gas/dust reservoirs needed to form further generations of stars.  This is in tension with most proposed scenarios for the origin of MPs.
\item We graphically summarise the comparison between models and predictions in Fig.~\ref{fig:truth_table}.
\end{enumerate}
\end{summary}

\section{Peculiar clusters: Fe spreads, CNO, and S-process Variations}\label{SEC:PECULIAR_GCS}

While large variations in light element abundances are almost universal among old and massive clusters, the abundances of heavier $\alpha$ (Si, Ca, Ti), Fe-peak (Fe, Ni) and n-capture (Sr, Ba, La, Eu)  elements within GCs vary little from star-to-star.

\subsection{Clusters with multimodal metallicity distributions: $\omega$~Cen, M54 and Terzan~5}
\label{sec:omega}

Understanding the formation and evolution of $\omega$~Cen, the most massive cluster in the Galaxy, represents a challenge for all the MP scenarios.
The presence of a wide metallicity range \citep[--2.2 $\leq \lbrack $Fe/H$ \rbrack \leq$--0.6 dex; e.g.][]{Johnson:10}  in its stars demands that it was massive enough to retain SN ejecta at very high velocity (or to accrete gas from its surroundings for long periods), allowing for multiple bursts of star formation, with each generation becoming progressively enriched in Fe. This possibly indicates that $\omega$~Cen constitutes the remnant of a tidally disrupted dwarf galaxy  \citep[e.g.,][]{Bekki:03}.
Although the observational scenario appears far more complex than for normal GCs, $\omega$~Cen also displays the key chemical signatures of MPs. Each metallicity subpopulation in $\omega$~Cen shows its own Na-O anti-correlation (with the possible exception of the most metal-poor stars), with the more metal-rich, He-rich stars \citep[Y $\geq$ 0.35;][]{Joo:13} showing a Na-O correlation \citep{Marino:11Omega}.
The extension of the Na-O anti-correlation is also more extended towards higher metallicity and the fraction of stars with high- and intermediate Na also increases with metallicity \citep{Marino:11Omega}. This is difficult to explain within the AGB scenario framework, as  the cooling flow from massive, metal-rich AGB stars would need to be delayed and further enriched by core-collapse supernovae to account for more extended Na-O anti-correlation towards higher metallicity  \citep[][]{Dantona:11Omega}. 

An increase in CNO sum and in the s-process element with [Fe/H] is also observed \citep{Marino:11Omega,Johnson:10}.
 Low-mass AGB stars (M $<$ 3 $\msun$) are observationally confirmed sites for s-process production, but they evolve on timescales longer (on the order of a Gyr) than the lifetimes of higher-mass AGB stars invoked to be responsible for the Na-O anti-correlation ($\sim100-200$ Myr, in the AGB scenario, see \S~\ref{sec:models_agb}).
Also, while AGB stars with masses $\lesssim$ 3 $\msun$ can produce Na, enhance the C+N+O content, and produce s-process elements, they cannot  deplete O or produce He. The most recent age estimates reports a maximum relative age spread of only $\sim$500 Myr among $\omega$~Cen populations \citep[][]{Tailo:16age}. 
Therefore low-mass AGBs, that evolve on longer timescales, cannot be responsible for the C+N+O and s-process pattern observed in $\omega$~Cen.

M~54 is the nearest extragalactic GC we can observe and the second most massive GC in the halo.
 Even though the M~54 metallicity distribution has a significantly smaller dispersion than $\omega$~Cen \citep[e.g.][]{Carretta:10M54}, both clusters have been proposed to represent a snapshot of nuclear star clusters in different stages of evolution. In the case of M~54, the associated dwarf galaxy, i.e. the Sagittarius is still visible, whereas the parent system once hosting $\omega$~Cen has been disrupted. 
Both metallicity groups in M~54 display their own Na-O anti-correlation, 
with the metal-poor group showing a less extended Na-O anti-correlation with respect to the metal-rich stars, as observed for $\omega$~Cen \citep[][]{Carretta:10M54}. 

Terzan~5 is a massive ($\sim$ 10$^{6}$ $\msun$) stellar system located in the bulge of the Galaxy. The two distinct red clumps in its CMD  \citep{Ferraro:09} have been linked to stellar populations with different metallicity \citep[although see][for an alternative explanation]{Lee:15}. Indeed, a large and multimodal metallicity distribution (--0.8 $\leq$ [Fe/H] $\leq$ +0.3 dex) has been reported \citep{Massari:14}, however there is no consensus on the presence of light element spreads in Terzan~5 \citep[e.g.][]{Origlia:11,Schiavon:17Ter5}. The $\alpha$-element abundance pattern of the metallicity sub-populations mirrors what is observed for field stars in the Bulge, with $\alpha$ enhancement up to about solar metallicities and a decreasing [$\alpha$/Fe] toward the solar ratio at super-solar [Fe/H] \citep{Origlia:11}. The presence of two distinct MSTOs suggests that the dominant sub-solar metallicity components developed $\sim$12 Gyr ago, while the super-solar groups formed only $\sim$ 4.5 Gyr ago after a prolonged period of quiescence \citep[e.g.][]{Ferraro:16}.  This finding has lead to the suggestion that Terzan~5 may constitute the remnant core of a dwarf galaxy, or perhaps even a surviving fragment of the formation of the original bulge. 

\subsection{Clusters with small unimodal Fe spreads and s-process bimodality}\label{sec:SCLUSTERS}	
 GCs characterised by a dispersion in their s-capture elements (e.g., M~22, NGC~1851, M~2, NGC~362, M~19, NGC~5286) have received a growing attention during the last years.
The observed s-process bimodal distribution is associated 
 with a split SGB in optical colours \citep[e.g.][]{Piotto:12} and, when C, N, and O abundances for unevolved stars are available, to variations in the net C+N+O content \citep[e.g.][]{YongCNO}. Each s-process group displays its own Na-O anti-correlation, with the average Na abundance positively correlated with s-process enrichment \citep[][]{Marino:11M22,Yong:14M2}. Finally, s-rich stars are possibly slightly enhanced in Fe \citep[e.g.,][]{DaCosta:15,Lim:17}. 

Since the presence of a [Fe/H] constrains the potential well in which a stellar system formed, a dispersion in [Fe/H] implies that the system was able to retain SN ejecta to host multiple star formation events\footnote{ The average [Fe/H] dispersions for MW GCs appear to be significantly smaller than the spectroscopic [Fe/H] spreads of $\sim$0.3 dex or more of dwarf galaxies, as no GC less luminous than M$_{\rm V}$ = --10 shows a substantial ($\geq$0.1 dex) [Fe/H] dispersions \citep[][]{Willman:12}.}. Indeed, it has been speculated that they 
represent the nuclear remnants of a tidally disrupted dwarf galaxy \citep[e.g.][]{DaCosta:15,Marino:15N5286}. 
This leads to the idea that GCs with small Fe variations would have contributed with a significant fraction of stars to the construction of the Galactic Halo, along with their host galaxies.

However, the presence of such small intrinsic Fe variations in a number of GCs is still debated, as they 
can be artificially introduced by the method used to derive atmospheric parameters of stars (\citealp{MucciarelliM22,LardoM2}; but see also \citealp{Lee:16}). 
For example, very little star-to-star Fe variation is measured when metallicity is measured from Fe II lines and the surface gravities are from photometry.  Conversely, when gravities are derived by imposing the ionisation equilibrium between the FeI and FeII, the [Fe/H] distribution is broad. Yet, the stellar gravities required to match [FeI/H] and [FeII/H] would lead to stellar masses for giants which are not physical \citep[e.g.][]{MucciarelliM22}. Interestingly, different FeI and FeII metallicity distributions are only observed in clusters which also show s-process and light-element variations. While the cause of the observed discrepancy between Fe abundances as inferred from Fe I and Fe II has been not yet determined, this finding suggests caution when measuring abundances using the classical spectroscopic approach on clusters with s-process variations.

Finally, the discrepancy between Fe abundances measured from Fe I and Fe II lines, which is observed for RGB stars with different s-processes in a few clusters is observed also in GC with no intrinsic variations in heavy elements in the AGB phase, where Fe I lines provide systematically lower abundances than RGBs \citep[e.g.][Wang et al., submitted]{Lapenna:15M62}. Currently, there is not explanation for this effect.

\subsection{The Blue Tilt in Cluster Populations} 
\label{sec:blue_tilt}

Observations of GC populations, especially around massive early type galaxies (ETGs) which contain thousands of such clusters, have shown that the metal poor population of clusters (i.e. the blue GCs) displays an average trend of becoming redder (more metal rich) as a function of increasing brightness \citep[e.g.,][]{Harris:09bluetilt}.  The origin of this ``blue tilt" is still uncertain, but a popular explanation for the phenomenon is that more massive clusters are able to retain not just the stellar ejecta (i.e., see \S~\ref{sec:models}) but also the SNe ejecta from a first generation of stars, and subsequently form a more metal rich second generation. The average metallicity of the cluster would then increases with each successive generation \citep[see][]{Strader:08,Bailin:09}.  One problem with such scenarios is that it is unclear how a cluster could retain the ejecta from SNe.

An alternative explanation, that can also account for the fact that the blue tilt is not observed in all GC populations, is that it is due to how the metal poor GC population is assembled, namely through the accretion of relatively low mass metal poor dwarf galaxies and their GC populations.  As lower mass dwarf galaxies have lower ISM pressures than their higher-mass counterparts, they are expected to form fewer high-mass clusters \cite[e.g.,][]{Kruijssen:15}.  Massive GCs will preferentially come from higher mass dwarf galaxies, which in turn are more likely to be metal rich.  This will result in a (statistical) upper envelope in the mass-metallicity plane for GC populations, skewing the mean metallicity to higher values for high cluster masses (Usher et al. in prep).  Such a scenario can be tested with the next generation of galaxy formation simulations that include GC formation and evolution \citep[e.g.,][]{Pfeffer:18}.

\section{Young Massive Clusters and Their Relation to Globular Clusters}
\label{sec:ymcs_gcs}

While historically  GCs were treated as objects that exclusively formed in the early Universe, it is now clear that objects with properties that are very similar to those expected of young GCs are still forming today.  Some of these YMCs have masses and densities well in excess of present day GCs, and their ages range from forming today to $\sim6-8$~Gyr.  While such clusters do exist in the Galaxy (with masses up to $\sim10^5$~\msun), they are difficult to study due to the often extreme (differential) extinction and crowding in the Galactic plane.  However, we are fortunate that our nearest extragalactic companions, the LMC and SMC host large populations of such clusters.  They are near enough that we can resolve them into their individual stars, especially with HST, and in some cases can obtain high-resolution spectra of individual stars.

A major result in the field in the past decade has been the finding that many of these clusters are not well represented by a single stellar isochrone, but instead show features such as dual MSs and extended MSTOs among other unexpected features.  The hope has been that these features are related to the MPs observed in the ancient GCs, and that they could then be used to pinpoint the physical mechanisms responsible for MPs.

\subsection{Extended Main Sequence Turn-offs in Young and Intermediate Age Clusters}
\label{sec:eMSTO}

The high precision photometry achievable with the Advanced Camera for Surveys (ACS) on HST allowed the construction of CMDs of massive young and intermediate age clusters in the LMC/SMC in unparalleled detail.  As is often the case, this increase in detail led to unexpected features that could not be explained within a traditional framework.  In this case, it was the discovery of eMSTOs in the intermediate age clusters ($1-2$~Gyr) in the LMC/SMC that could not be explained by photometric uncertainties or stellar binarity.  This was first reported by \citet{Bertelli:03} and \citet{Mackey:07} and shown to be a general characteristic in subsequent works \citep[e.g.,][]{Mackey:08,Milone:09,Piatti:14}.

\begin{marginnote}[]
\entry{eMSTO}{Extended Main Sequence Turn-off}
\end{marginnote}

The initial explanation for the eMSTOs was that the clusters were formed in an extended star-forming event, lasting $200-700$~Myr \citep[e.g.,][]{Milone:09,Goudfrooij:14}.   Due to this possibility, many works have attempted to link the observations of the eMSTO clusters with those of the ancient GCs hosting MPs \citep[e.g.,][]{Goudfrooij:14}.  However, subsequent work has shown that the eMSTO phenomenon is unlikely to be caused by an actual age spread within the clusters (see \S~\ref{sec:ymcs}).  Subsequent studies have found that YMCs with ages between $20-300$~Myr also show eMSTOs, and that the inferred age spread was directly proportional to the age of the cluster \citet{Niederhofer:15rotation}.  Additionally,  studies focused on other regions of the CMDs that should also be affected by age spreads have not been found to be in agreement with the age-spread interpretation \citep[e.g.,][]{Li:16}.   Finally, at $\sim2$~Gyr, NGC~1978 does not show an eMSTO \citep[][]{Martocchia:18b} despite its relatively high mass.

This points instead toward a stellar evolutionary affect.  One such affect is stellar rotation, first proposed by \citet{Bastian:09rotation} and subsequently studied in more detail in \citet{Brandt:15} using the Geneva stellar evolutionary models that include rotation. Such models do well in predicting the relation between the inferred age spread an the age of the cluster, as well as the lack of eMSTOs in clusters with ages above $\sim2$~Gyr due to magnetic breaking of the stars.

Finally, recent high-resolution studies of A and F ($1-2.5$\msun) stars have found evidence for light-element abundance (Na, O, Mg) spreads in rapidly rotating stars in open clusters \citep{Pancino:18}.  The origin of these variations (and their link to GCs) is still unknown, but rotational mixing and diffusion are possible causes.

It is striking that the eMSTO phenomenon disappears at (nearly) the same age that MPs on the RGB begin to be seen \citep{Martocchia:18a,Martocchia:18b}.  How/whether these two phenomenon are related is a rich avenue for future work.  

\subsection{Split Main Sequences}

Another surprising feature that has been found in  resolved CMDs of YMCs in the LMC/SMC was that many of them, when viewed in the blue/UV filters displayed bi-modal (i.e. split) MSs \citep{Milone:15n1856}.  At first glance, this appears to be similar to the split MS in ancient GCs which are due to light element abundance spreads (e.g., He, C, N, O spreads).  However, \citet{Milone:15n1856} investigated possible causes of the splits, creating stellar models that included the abundance spreads, iron spreads, C+N+O spreads and also age spreads.  They conclude that none of the models were able to explain the split MS observed in clusters like NGC~1856 ($\sim300$~Myr, $\sim10^5$~\msun).

\citet{Dantona:15} used the SYCLIST stellar models \citep{Georgy:14} that include rotation (including inclination effects) to model NGC~1856, and showed that rotation could explain the observed MS split if the stellar rotation distribution was bi-modal with a minor peak peak at $\omega<0.3$ and a dominant peak at $\omega\sim0.9$.  It is interesting to note that in all the YMCs in the LMC studied to date with split MS, the red MS (corresponding to the rapid rotators) is generally the dominant population (between 42\% and 75\% - e.g., \citealt{Milone:16n1755,Milone:17n1866}).  These stars would be rotating much faster than typically found in the field or in lower mass open clusters \citep{McSwain:05}.

Such an extreme rotational distribution should lead to observationally detectable signatures, as a large population of rapid rotators should have a high rate of Be stars, i.e. stars near the critical rotation limit with partially ionised decretion discs.  \citet{Bastian:17} looked for such a population of Be stars and indeed found a much higher fraction in the $\sim100$~Myr cluster NGC~1850 and the $\sim300$~Myr cluster NGC 1856.  In both clusters, the authors found Be fractions between $30-60$\% near the MSTO, much higher than found in the field or in lower mass clusters.  These observations confirmed the high fraction of rapid rotators in YMCs, lending support to the idea that the split MS is caused by a bi-modal rotational distribution.

However, further observations to measure the actual rotational distribution in YMCs are required to directly test this scenario.  Preliminary results appear to confirm the bi-modal rotational distribution with a large fraction of rapidly rotating stars \citep{Dupree:17}.  If true, the conclusion would be that stars forming in dense/massive clusters would retain a signature of their origin, namely in their rapid rotation rates.  Although why stars born in clusters would preferentially be born with high rotation rates is currently unknown.

\subsection{Chemical Anomalies in YMCs?}

While YMCs have provided strong tests for the theories of the formation of MPs, it is not yet clear whether they host such abundance anomalies. As discussed in \S~\ref{sec:age_mass} initial spectroscopic studies of a limited number stars in massive young and intermediate age clusters in the LMC did not find evidence of MPs \citep{Mucciarelli:08,Mucciarelli:14LMC}.  This has been confirmed through photometric studies based on large samples \citep{Martocchia:17N419,Martocchia:18a}.

\begin{marginnote}[]
\entry{RSG}{Red Supergiant Star}
\end{marginnote}

The young and intermediate age LMC and SMC clusters are quite massive, relative to their open cluster counterparts in the Galaxy, however as discussed in \S~\ref{sec:ymcs}, YMCs with much higher masses (by factors of 10 to 1000) are known to exist.  However, the distances to these extragalactic objects generally makes it impossible to obtain high precision photometry or spectroscopy for individual stars.  Hence, some studies have attempted to search for the spectroscopic fingerprint of MPs in integrated light.  \citet{CabreraZiri:16} and \citet{Lardo:17Antennae} have exploited the fact that YMCs are dominated by the light of RSGs at young ages (in the near-IR), and that RSGs all have similar temperatures, meaning that their integrated light can be studied as a single RSG.  If MPs would be present in these massive YMCs, we would expect that their Al and Na abundances would be higher than that of field RSGs at the same Fe-abundance.  These authors studied four clusters with masses between $5-20 \times10^5$~\msun, and searched for evidence of Al enhancement, although none was found in any of the clusters, despite their high masses. This RGB focussed technique is sensitive to chemical anomalies in stars above $\sim15$~\msun \citep[e.g.,][]{Davies:08}, although integrated light spectroscopy can in principle be used to search for MPs at any age, with proper modelling of its stellar populations (e.g., Hernandez et al.~2017).

One potential caveat to note about the previous studies is that they are not comparing like-with-like, at least in terms of stellar mass.  All studies of young and intermediate age clusters have focussed on the evolved portions of the CMD (e.g., the RGB), which at 200~Myr or 2~Gyr corresponds to a stellar mass of $\sim3.6$~\msun\ and $\sim1.5$~\msun, respectively (at [$Fe/H]=-0.7$).  At ages of $6$ and $10$~Gyr the stellar mass on the RGB is $\sim1.0$~\msun\ and $\sim0.9$~\msun, respectively.  While the main sequence for the LMC/SMC young/intermediate massive clusters is out of range for spectroscopy with existing instruments, there is potential to use HST to obtain N-sensitive photometry to compare the same mass range in young and ancient clusters (i.e., $< 0.8$~\msun).  Additionally, future instruments like JWST or the E-ELT may provide important insights at lower stellar masses.

\section{Multiple Populations on Galaxy Scales}
\label{sec:mps_galaxy_scales}

Dwarf galaxies have stellar masses ranging from the GC mass scale up to a few $\times10^9$~\msun. In many cases, their stellar populations are not too dissimilar from that of certain GCs (like $\omega$-Cen and M54), with modest metallicity spreads and a dominant old stellar population (see~\S~\ref{sec:omega}). It is normally assumed that MPs are not present in the field stars in dwarfs, due to 1) the assumption that MPs are restricted to GCs and 2) the low fraction ($\sim3$\%) of 2P stars in the field of the MW halo \citep[e.g.,][]{Martell:11} which is thought to come from, at least partially, from accreted satellite dwarf galaxies. We can infer a lack of a large population of stars with large $\Delta(Y)$ values within local dwarf galaxies, based on the morphology of the HB.  The HB of dwarf galaxies lack to the ``extreme" stars seen in GCs with large Y spreads (e.g., NGC~2808).  For example, detailed modelling of the HB of the Carina Dwarf galaxy did not lead to evidence of Y spreads within the populations (although age and Fe spreads were identified - \citealt{Savino:15}).  Additionally, \citet{Norris:17} searched for MPs in the Carina dwarf galaxies in 63 RGB stars (looking for an Na-O spread) and only found stars with typical abundance patterns, i.e. 1P stars.

Stepping further afield, \citet{Strader:13} studied a very massive ($\sim2\times10^8$~\msun) and dense (R$_{\rm h}=24$~pc) ultra-compact dwarf galaxy around the Virgo elliptical galaxy, M60 (M60-UCD1).  The authors find evidence for the object to be enriched in N ([N/Fe]$=+0.61$) and Na ([Na/Fe]$=+0.42$), hence it likely hosts MPs, with a large population of highly enriched 2P stars.

While studies of MPs and chemical anomalies have largely focussed on massive and dense star clusters, there is growing evidence that they may be present outside clusters, making up a significant fraction of the stars in certain parts of galaxies.  \citet{Schiavon:17bulge} discovered a large population of N-rich stars, which display correlations between [N/Fe] and [Al/Fe], as well as being anti-correlated with [C/Fe], i.e. they display the same chemical anomalies as stars in GCs.  The authors focussed on the low metallicity regime and found that  for $[Fe/H] <  -1$, the chemically anomalous stars make up $\sim7$\% of the stars of the bulge/inner-halo.  Extrapolating their results to the full bulge/inner halo, they estimate that the mass of enriched stars is a few times $10^8$~\msun, which is a factor of $\sim8$ more than the mass of the entire Galactic GC system.  This fact, and the lack of correspondence between the enriched star and GC population metallicity distributions, suggests that the discovered enriched stars in the bulge/inner-halo did not originate from dissolved GCs.

If true, this would suggest that MPs may not be a product of only GCs, but may instead be a general feature of certain stellar populations.  While currently still inconclusive, there is tantalising evidence that MPs may be present in other dense and old stellar populations.  For example, the mean [N/Fe] and [Na/Fe] abundances of ETGs increase with increasing velocity dispersion \citep[e.g.,][]{Schiavon:07,Conroy:14} which could imply that the fraction of enriched stars is an increasing function of velocity dispersion.  Recently, \citet{vanDokkum:17} have used high S/N spatially resolved spectra of massive ETGs and find that the mean [Na/Fe] abundance increases towards the galaxy centres while [O/Fe] decreases, again suggesting that MPs may be present in the centres of such systems.  While high velocity dispersion within ETGs is also positively correlated with high [Mg/Fe] \citep[e.g.][]{Walcher:15}, \citet{vanDokkum:17} found that relative to the outskirts of the galaxies, [Mg/Fe] was depressed in the central regions.  Hence, the centres of ETGs appear to show many of the trends seen in MPs.

Another potential link between MPs and the massive ETGs is through the UV-upturn \citep[e.g.,][]{Oconnell:99}.  The origin of the UV-upturn is still under debate but the presence of a large number of extreme HB stars is one of the leading contenders.  As seen in Galactic GCs, like NGC~2808, the presence of a large He spread amongst cluster stars is correlated with an extreme population of HB stars (metallicity also effects the fraction of stars that pass through an extreme HB period).  Hence, if ETGs do host MPs, it would imply that the UV-upturn is caused by large He spreads, which would be correlated with large Na and N-spreads \citep{Chantereau:18}.

Further work is needed to explicitly test if MPs are present within ETGs and if so, in what fractions.  However, if MPs are found to make up a significant fraction of ETG stars, it would have a dramatic effect on our understanding of MPs and their origin.  It may imply, for example, that we need to explore non-cluster focussed scenarios for the origin of MPs.

\section{Future Directions}

Throughout this review we have attempted to highlight topics that are particularly uncertain and which new theoretical and observational studies are likely to lead to important advances.  Here we briefly summarise some of the directions that we feel are likely to be the most fruitful in the next few years.

\begin{itemize}
\item While observations of evolved stars in YMCs have not revealed the presence of MPs, it is not clear if MPs are absent or restricted in the stellar mass range where they can appear.  The unexpected transition at $\sim2$~Gyr, below which MPs are not found in evolved stars and above which they are, suggests that MPs may be present in many YMCs, but only in low mass stars  (i.e., lower-mass main sequence stars).
\item In order to identify the cluster parameter(s) that control whether MPs are present (age, mass, density, metallicity, etc) the parameter space of clusters should be further sampled.  Looking at low-density GCs in the outer Galactic halo, or those that have been accreted could be particularly fruitful.  Also, extending the age range of clusters under study may place stricter limits on the appearance of MPs.
\item Further work quantifying how the properties of MPs within clusters depend on the cluster properties would be very beneficial.  Is cluster mass or density the controlling factor for the fraction of enriched stars or the degree of abundance spreads within clusters?
\item To date, only a handful of GC stars have been fully characterised in terms of their abundances (He, C, N, O, Na, Al, Mg, etc). Systematic studies of the precise way all these elements are related, and of the variety between clusters may help pinpoint the origin of MPs.  Dissecting the (pseudo)colour-colour diagrams of the HST UV GC survey may offer an efficient means to search many of these correlations.  What causes the spread in the 1P stars in the pseudo-colour diagrams in some clusters and not in others?  Detailed modelling of the colour spreads in is needed to characterise the abundance variations in a large sample of GCs (as well as confirmation through spectroscopic follow-up).
 If spectroscopy confirms that the colour spread among 1P stars is due to He variations (associated with small-or-no C-N-Na-O variations), alternative physical mechanisms for the origin of MPs --other than stellar nucleosynthesis-- will need to be investigated.
\item As discussed in \S~\ref{sec:mps_galaxy_scales} there is tentative evidence that MPs may not be restricted to GCs but may be present in other environments as well (dwarf galaxies, bulge/inner halos of galaxies and ETGs).  Studies confirming or refuting this may result in a major breakthrough in the field.
\item Recent theoretical studies have largely focussed on developing existing scenarios, exploring ways in which the models can be changed in order to provide a better match to observations.  We argue that the present observations do not support the traditional theories of self-enrichment through the formation of multiple generations of stars.  Hence, new theories for the origin of MPs (e.g., non-standard stellar evolution, very massive stars, etc) should be encouraged and developed to test against the wealth of observational data now in hand.
\item One property of stars that affects stellar evolution, which is dependent on environment, is stellar rotation.  Stars in dense/massive young clusters rotate significantly faster than those in the field or lower mass open clusters.  Additionally, the age boundary for whether MPs are present ($2-2.5$~Gyr) in evolved stars is also the boundary ($\sim1.5-1.6$~\msun) at which MSTO and RGB stars would be magnetically braked (i.e., at this age clusters no longer show extended main sequence turn-offs).  Could MPs be caused by a non-standard stellar evolutionary effect linked to rotation?
\end{itemize}

\section*{DISCLOSURE STATEMENT}
The authors are not aware of any affiliations, memberships, funding, or financial holdings that
might be perceived as affecting the objectivity of this review. 

\section*{ACKNOWLEDGMENTS}
We are grateful to Soeren Larsen,  Ivan Cabrera-Ziri, Corinne Charbonnel, Maurizio Salaris, Mark Gieles, Emanuele Dalessandro, Alessio Mucciarelli, Chris Usher,  William Chantereau, Elena Pancino, Santi Cassisi, Francesca D'Antona, Henry Lamers, Eugenio Carretta, and Angela Bragaglia for helpful discussions and detailed comments on earlier versions of the manuscript.  Additionally, we thank the editor, Sandy Faber, for suggestions that significantly improved the manuscript.   NB  acknowledges financial support from the Royal Society (University Research Fellowship) and the European Research Council (ERC-CoG-646928, Multi-Pop).  CL thanks the Swiss National Science Foundation for supporting this research through the Ambizione grant number PZ00P2\_168065.

%

\bibliographystyle{ar-style2}
\bibliography{biblio}

\end{document}